\documentclass[a4paper,12pt]{article}

\usepackage{epsfig}
\usepackage{color}
\usepackage{amssymb}
\usepackage{amsmath}
\usepackage{amsfonts}
\usepackage[bookmarks=false]{hyperref}
\usepackage{fullpage}

\begin{document}
\newcommand{\newc}{\newcommand}
\newc{\ba}{\begin{array}}
\newc{\ea}{\end{array}}
\newc{\bea}{\begin{eqnarray}}
\newc{\eea}{\end{eqnarray}}
\newc{\beastar}{\begin{eqnarray*}}
\newc{\eeastar}{\end{eqnarray*}}
\newc{\beq}{\begin{equation}}
\newc{\eeq}{\end{equation}}
\newc{\bestar}{\begin{equation*}}
\newc{\eestar}{\end{equation*}}
\newc{\ben}{\begin{enumerate}}
\newc{\een}{\end{enumerate}}
\newc{\bi}{\begin{itemize}}
\newc{\ei}{\end{itemize}}
\newc{\mymed}{\vspace{0.01cm}}

\newcommand{\beal}{\begin{aligned}}
\newcommand{\eeal}{\end{aligned}}
\renewcommand{\thefootnote}{\fnsymbol{footnote}}
\renewcommand{\Re}{\widetilde{\operatorname{\mathfrak{Re}}}}
\renewcommand{\Im}{\operatorname{\mathfrak{Im}}}
\newcommand{\pslash} {\ensuremath{\slash \hspace*{-2mm} p}}
\newc{\lam}{\lambda}
\newc{\lamp}{\lambda^\prime}
\newc{\lampp}{\lambda^{\prime\prime}}
\newc{\Lam}{\Lambda}
\newc{\BLam}{{\mathbf{\Lam}}}
\newc{\eps}{\epsilon}
\newc{\kap}{\kappa}

\newc{\ra}{\rightarrow}
\newc{\ovl}{\overline}
\newc{\Tr}{{~\rm Tr}}

\newc{\itRahmen}[2]{
\begin{center}\fbox{\parbox{#1 cm}{\it #2}}
\end{center}}

\newc{\del}{\partial}
\newc{\veva}{\langle H_1\rangle}
\newc{\vevb}{\langle H_2\rangle}
\newc{\onehalf}{\textstyle \frac{1}{2} \displaystyle}
\newc{\onethird}{\textstyle \frac{1}{3} \displaystyle}
\newc{\mzero}{m_0}
\newc{\mhalf}{{m_{1/2}}}
\newc{\tanb}{\tan\beta}
\newc{\Psix}{{\mathrm{P}_{\!6}}}
\newc{\nPsix}{{\not\!\Psix}}
\newc{\nPsixU}{{\s{\mathrm P}_{\!6}}}

\newc{\muon}{\mu}
\newc{\azero}{A_0}
\newc{\neutralino}{\tilde\chi^0}
\newc{\selectron}{\tilde e}
\newc{\smuon}{\tilde\mu}
\newc{\sneu}{\tilde\nu}
\newc{\higgs}{h^0}
\newc{\sgnmu}{\textrm{sgn}(\mu)}
\newc{\gev}{\mbox{~GeV}}
\newc{\tev}{\mbox{~TeV}}
\newc{\mgut}{{M_{GUT}}}
\newc{\mweak}{{M_{W}}}

\newbox\charbox
\newbox\slabox
\def\s#1{{      
    \setbox\charbox=\hbox{$#1$}
    \setbox\slabox=\hbox{$/$}
    \dimen\charbox=\ht\slabox
    \advance\dimen\charbox by -\dp\slabox
    \advance\dimen\charbox by -\ht\charbox
    \advance\dimen\charbox by \dp\charbox
    \divide\dimen\charbox by 2
    \raise-\dimen\charbox\hbox to \wd\charbox{\hss/\hss}
    \llap{$#1$}
}}

\newc{\ssup}{\tilde{u}}
\newc{\ssdown}{\tilde{d}}
\newc{\ssstrange}{\tilde{s}}
\newc{\sscharm}{\tilde{c}}
\newc{\sstop}{\tilde{t}}
\newc{\ssbottom}{\tilde{b}}
\newc{\sse}{\tilde{e}}
\newc{\ssmu}{\tilde{\mu}}
\newc{\sstau}{\tilde{\tau}}
\newc{\ssnue}{\tilde{\nu}_{e}}
\newc{\ssnumu}{{\tilde{\nu}_{\mu}}}
\newc{\ssnutau}{{\tilde{\nu}_{\tau}}}
\newc{\ssbnue}{\bar{\tilde{\nu}}_{e}}
\newc{\ssbnumu}{\bar{\tilde{\nu}}_{\mu}}
\newc{\ssbnutau}{\bar{\tilde{\nu}}_{\tau}}
\newc{\charge}{{\chi}}
\newc{\Higgs}{H^0}
\newc{\Azero}{A_0}

\newc{\nue}{\nu_e}
\newc{\numu}{\nu_{\mu}}
\newc{\nutau}{\nu_{\tau}}
\newc{\bnue}{\bar{\nu_e}}
\newc{\bnumu}{\bar{\nu_{\mu}}}
\newc{\bnutau}{\bar{\nu_{\tau}}}
\newc{\Br}{\mathrm{Br}}
\setlength{\parindent}{1.5em}
\setlength{\parskip}{0.5ex plus 0.5ex minus 0.2ex}
\setcounter{secnumdepth}{5}
\setcounter{tocdepth}{5}

\newc{\stext}[1]{{\color{red}  #1}}

\newcommand {\dmw} {\frac{\delta \mw^2}{\mw^2}}
\newcommand {\dmz} {\frac{\delta \mz^2}{\mz^2}}
\newcommand {\cbeta} {\cos_{\beta}}
\newcommand {\sbeta} {\sin_{\beta}}
\newcommand {\lneut} {\tilde\chi_1^0}

\begin{center}

\vspace{1cm}

{\Large {\bf Radiative~Corrections~to~the~Neutralino~Dark~Matter }}

{\Large {\bf Relic Density - an Effective Coupling Approach}}

\vspace{1cm}

{\sc Arindam Chatterjee}$^{a,\,}$\footnote{arindam@th.physik.uni-bonn.de},
{\sc Manuel Drees}$^{a,b\,}$\footnote{drees@th.physik.uni-bonn.de} and 
{\sc Suchita Kulkarni}$^{c,\,}$\footnote{suchita.kulkarni@lpsc.in2p3.fr} \\

\vskip 0.15in
{\it
$^a${Physikalisches Institut and Bethe Center for Theoretical Physics, 
\\ Universit\"at Bonn, D-53115 Bonn, Germany\\
$^b$KIAS, School of Physics, Seoul 130--722, Korea\\
$^c$LPSC, 53 Av. des Martyrs, 38000, Grenoble, France}
}

\vskip 0.5in

\abstract{In the framework of the minimal cosmological standard model,
  the $\Lambda$CDM model, the Dark Matter density is now known with an
  error of a few percent; this error is expected to shrink even
  further once PLANCK data are analyzed. Matching this precision by
  theoretical calculations implies that at least leading radiative
  corrections to the annihilation cross section of the dark matter
  particles have to be included. Here we compute one kind of large
  corrections in the context of the minimal supersymmetric extension
  of the Standard Model: corrections associated with two--point
  function corrections on chargino and neutralino (collectively
  denoted by $\tilde \chi$) lines. These can be described by effective
  $\tilde \chi$--fermion--sfermion and $\tilde \chi$--$\tilde
  \chi$--Higgs couplings. We also employ one--loop corrected
  $\tilde\chi$ masses, using a recently developed version of the
  on--shell renormalization scheme. The resulting correction to the
  predicted Dark Matter density depends strongly on parameter space,
  but can easily reach 3\%.}

\end{center}
\pagenumbering{arabic}
\clearpage
\pagestyle{plain}
  
\section{Introduction}

The presence of Dark Matter is by now well established. In fact,
combining data from several cosmological observations, including in
particular analyses of cosmic microwave background anisotropies,
allows to determine the universally averaged Dark Matter density with
an error of about 3\% \cite{Komatsu:2010fb}, assuming that the standard
cosmological ``$\Lambda$CDM'' model, which combines a cosmological constant
with essentially non--interacting cold Dark Matter, is correct. Note
that this simple model describes all cosmological data satisfactorily.
This error is expected to shrink to the 1\% range once the PLANCK
collaboration \cite{Planck} releases their data.

A particularly attractive class of particle physics candidates for Dark Matter
are Weakly Interacting Massive Particles (WIMPs). Their relic density can be
calculated precisely in the framework of standard cosmology \cite{kotu},
assuming that the reheat temperature after inflation was higher than about 5\%
of the WIMP mass \cite{Giudice:2000ex,Fornengo:2002db,kaki}. Input in this
calculation is the WIMP annihilation cross section into Standard Model (SM)
particles; the predicted relic density scales essentially like the inverse of
this cross section. A precise prediction for the relic density therefore
requires an accurate calculation of the total WIMP annihilation cross section.

This can only be accomplished within a specific WIMP model. Since the
SM doesn't contain a WIMP (nor any other plausible candidate for cold
Dark Matter), it needs to be extended. The best motivated WIMPs are
those that arise in extensions of the SM that have {\em independent}
motivation, not related to Dark Matter.

Here we focus on supersymmetry, more specifically, on the minimal
supersymmetric extension of the SM (MSSM) \cite{Drees2}. Superpartners of all
SM particles not only stabilize the weak hierarchy against radiative
corrections, they also allow single step Grand Unification of all gauge
interactions. We assume that $R-$parity is exact, so that the lightest
superparticle (LSP) is stable, thereby satisfying one of the conditions on a
successful WIMP candidate. Moreover, we assume that the supersymmetry breaking
terms conserve CP. Barring several finetunings, experimental constraints on CP
violation, mainly from electric dipole moment measurements, require their
phases to be small, unless superpartners of first (and second) generation
fermions are very heavy \cite{Ibrahim:1998je, Pospelov:2005pr}. However, we do
not assume a particular scenario for supersymmetry breaking (presumably
described by high--scale dynamics), instead simply treating the relevant
weak--scale soft breaking parameters as independent inputs.

The MSSM contains four potential WIMPs, the three sneutrinos $\tilde
\nu_i$ and the lightest neutralino $\tilde \chi_1^0$. However,
sneutrinos have full weak strength vector couplings to the $Z$,
leading to a cross section for scattering on protons that is well
above experimental constraints, unless their masses are so large that
their predicted relic density is well above the observed value. This
leaves the lightest neutralino as unique WIMP candidate in the MSSM
\cite{Falk:1994es}.

While there have been many analyses of $\lneut$ as WIMP candidate, nearly all
of them compute the relevant annihilation cross sections only at tree
level. However, in order to achieve percent level precision, at least the
leading radiative corrections need to be taken into account. These include QCD
corrections for strongly interacting final states \cite{djkn, Herrmann:2007ku,
  Freitas:2007sa, Herrmann:2009wk, Herrmann:2009mp}, but at least leading
electroweak corrections should also be included; with ``leading'' we mean
corrections that are parametrically larger than the naive one--loop factor
$\alpha/\pi$.

Of course, a complete one--loop calculation would be ideal, and has in fact
already been performed for a few points of parameter space \cite{Baro:2007em,
  Baro:2009na, Boudjema:2011ig}.  However, calculating cross sections at one
loop order involves computing a large number of diagrams. This not only needs
a large effort even in the days of largely automatized calculation of
one--loop amplitudes, the resulting expressions are also very complex, and
their evaluation is very CPU intensive. With present--day computers the
results of full one loop calculations can therefore not be used if the relic
density has to be evaluated for many different choices of parameters.

It is therefore useful to develop alternative ways to estimate the bulk of
electroweak radiative corrections to the relic density. One class of diagrams
that can lead to large corrections involves the exchange of a relatively light
boson between the annihilating WIMPs. These corrections can even become
non--perturbative if the ratio of the boson and WIMP masses is smaller than
the relevant squared coupling $\alpha$ \cite{Hisano:2003ec, Hisano:2004ds,
  ArkaniHamed:2008qn, Iengo:2009ni}, and they can remain significant even for
larger mass of the exchanged boson. In the context of the MSSM these
corrections are usually small \cite{Drees:2009gt}, except in regions of
parameter space where $\lneut$ is a heavy higgsino-- or wino--like state
\cite{Hryczuk:2010zi, Hryczuk:2011vi}.

In this article we are interested in another class of potentially
large radiative corrections, which can be described by running or
effective couplings. For example, it has been recognized some time ago
\cite{djkn} that running quark masses, or Yukawa couplings, evaluated
at a scale near $m_{\tilde \chi_1^0}$, should be used in the
calculation of the $\lneut$ relic density. Similarly, most
calculations of the $\lneut$ relic density employ running electroweak
couplings, based on the QED coupling at the weak scale (often, at
scale $M_Z$), rather than the on--shell value of $\alpha_{\rm em}$.

In the limit of exact supersymmetry these electroweak gauge couplings
not only appear in $\lneut$ annihilation into final states containing
gauge bosons, and in $s-$channel $Z-$exchange diagrams; they also
determine the couplings of $\lneut$ to a Higgs boson and a neutralino
or chargino, as well as the gauge contributions to the $\lneut$
couplings to a sfermion and a fermion, which have the structure of
Yukawa couplings. However, when supersymmetry is broken, SM particles
and their superpartners (for example, $q$ and $\tilde q$) no longer
have the same masses, which leads to different contributions to gauge
and gaugino two--point functions. In particular, for energies between
the squark and quark mass scales, the (s)quark superfields no longer
contribute to the running of the gauge Yukawa couplings, but the
quarks still contribute to the running of the true gauge couplings.
Corrections to the gauge Yukawa couplings are the supersymmetric
analog of the oblique corrections, and have thus been dubbed
``super--oblique'' corrections \cite{Cheng:1997sq,
  Katz:1998br}.\footnote{This apparent hard breaking of supersymmetry
  caused at one--loop order by soft breaking terms has first been
  noticed in the context of strong interactions in
  ref.\cite{Hikasa:1995bw}.}

While the first analyses of these super--oblique corrections only
included terms that grow logarithmically with the masses of the
heaviest superpartners \cite{Cheng:1997sq, Cheng:1997vy}, references
\cite{Nojiri:1997ma, Kiyoura:1998yt} include the full quark and squark
corrections to leptonic processes, thereby also including terms that
approach a constant for large squark mass. Finally, in
\cite{Guasch:2002ez} this formalism was extended to include {\em all}
(s)fermion loop corrections involving gauge Yukawa couplings to the
chargino and neutralino two--point functions. This allowed the
introduction of effective neutralino and chargino couplings to
sfermions and fermions; here we apply the same formalism also to the
gauge Yukawa couplings involving a Higgs boson and two neutralinos or
charginos. 

Note that these corrections are enhanced not only by the logarithm of
the ratio of the mass of the heaviest sparticle to the LSP mass, but
also by the large number of fields contributing with the same sign
(since the contributions to diagonal gaugino two--point functions go
like the square of the relevant charge or coupling). The latter factor
is more important, unless one is considering a scenario where some
sfermions are more than an order of magnitude heavier than the LSP. A
calculation \cite{dhx} of radiative corrections to $\tilde \chi_2^0
\rightarrow \tilde \chi_1^0 e^+e^-$ decays found that the bulk of the
corrections to the partial width could be described by these effective
couplings even in a scenario where all superparticles are relatively
light.

Since we include (certain) one--loop corrections to chargino and
neutralino two--point functions, we also use one--loop renormalized
neutralino and chargino masses. We use a recently developed
\cite{CDKX} variant of the on--shell renormalization scheme
\cite{Fritzsche:2002bi}, where corrections to physical masses are
small for (nearly) all combinations of parameters, in contrast to
earlier schemes, which lead to large corrections when the $SU(2)$
gaugino and higgsino mass parameters become close in magnitude. The
effect of this mass renormalization on the relic density is therefore
usually (although not always) small.

This paper is organized as follows: in section \ref{sec:Form} we outline our
formalism and discuss the structure of the corrections included in the
effective couplings. Section \ref{sec:Imp} contains technical details of our
numerical calculation. Section \ref{sec:Num} is devoted to the discussion of
numerical results and we conclude in section \ref{sec:Conc}. Details of the
renormalization scheme we chose are given in the Appendices.

\section{Formalism}
\label{sec:Form}

In this section we describe the formalism of effective couplings, closely
following \cite{Guasch:2002ez}, and discuss in some detail the nature of the
corrections treated in this formalism. Here we are interested in corrections
to the Yukawa couplings of the electroweak neutralinos and charginos. The
required renormalization of the chargino, neutralino, Higgs and electroweak
gauge sectors is described in the Appendices.

At tree level the neutralino and chargino couplings in question can be
written as \cite{Guasch:2002ez}
\begin{eqnarray}
{\cal L}_{\tilde \chi_\alpha^0 \tilde f f} &=& - \sqrt{2} g 
\overline{\tilde\chi_\alpha^0} \left[ \left(T^f_3 N^*_{\alpha 2} + N^*_{\alpha 1}\tan\theta_W 
\left( Q_f - T^f_3 \right)\right) P_L \tilde f^*_L - \tan\theta_W Q_f N_{\alpha
    1} P_R \tilde f^*_R \right] f \nonumber \\
&-& \frac {g m_f}{\sqrt{2} M_W f_f(\beta)} \overline{\tilde\chi_\alpha^0}
\left[ N^*_{\alpha h_f} P_L \tilde f^*_R + N_{\alpha h_f} P_R \tilde
  f^*_L \right] f  \ + \ h.c. \,;  \label{neutcoup} \\
{\cal L}_{\tilde \chi_i^+ \tilde d u} &=& - g \overline{\tilde
  \chi_i^+}U^*_{i1} P_L u \tilde d_L^* + \frac {g m_d} {\sqrt{2} M_W
  \cos\beta} \overline{\tilde\chi_i^+} U^*_{i2} P_L u \tilde d_R^* \nonumber\\
& + & \frac {g m_u}{\sqrt{2} M_W \sin\beta} \overline{\tilde \chi_i^+}
V_{i2} P_R u \tilde d_L^* \ + \ h.c. \,. \label{charcoup}
\end{eqnarray}
The unitary $4 \times 4$ matrix $N$ appearing in eq.(\ref{neutcoup})
diagonalizes the neutralino mass matrix, while the unitary $2 \times
2$ matrices $U, \, V$ appearing in eq.(\ref{charcoup}) diagonalize the
chargino mass matrix. Furthermore, $g$ is the $SU(2)$ gauge coupling,
$\theta_W$ is the weak mixing angle, $T^f_3$ and $Q_f$ are the third
component of the weak isospin and electric charge of fermion $f$,
respectively, $P_{R,L} = (1 \pm \gamma_5) / 2$ are the chiral
projectors, $m_f$ and $M_W$ are the masses of fermion $f$ and of the
$W$ boson, respectively, and $\tan\beta$ is the ratio of vacuum
expectation values (vevs) of the two neutral Higgs fields required in
the MSSM. Finally, the function $f_f(\beta)$ appearing in
eq.(\ref{neutcoup}) is $\sin\beta$ for up--type quarks, and
$\cos\beta$ for down--type quarks and charged leptons; similarly, the
index $h_f = 4 \ (3)$ for up--type quarks (down--type quarks and
charged leptons). The chargino couplings to a neutrino and a charged
slepton have the form of eq.(\ref{charcoup}) with $u \rightarrow \nu,
\ \tilde d \rightarrow \tilde \ell$ (and $m_u \rightarrow m_\nu = 0$
in our treatment). The chargino couplings to a down--type quark and an
up--type squark again have the form of eq.(\ref{charcoup}), with $ u
\leftrightarrow d, \ U \leftrightarrow V$ and $\tilde \chi_i^+
\leftrightarrow \left( \tilde \chi_i^+ \right)^C$, where the
superscript $C$ denotes charge conjugation.

Equations (\ref{neutcoup}) and (\ref{charcoup}) have been written for
neutralino and chargino mass eigenstates, but fermion and sfermion
current eigenstates. In our numerical treatment we have ignored flavor
mixing in both the fermion and sfermion sectors. Quark mixing is known
to be quite small, and neutrino mixing is irrelevant for our purposes
(since neutrinos remain very nearly degenerate and massless, compared
to the sparticle masses). Moreover, sfermion flavor mixing is strongly
constrained by experimental upper bounds on flavor changing neutral
currents. However, in our numerical treatment we include $\tilde f_L -
\tilde f_R$ mixing for third generation sfermions.

Note that the first line in eq.(\ref{neutcoup}) and the first term in
eq.(\ref{charcoup}) come from gauge interactions, whereas the second line in
eq.(\ref{neutcoup}) as well as the second and third term in
eq.(\ref{charcoup}), which are proportional to SM fermion masses, originate
from superpotential couplings. Technically the corrections we are interested
in originate from fermion--sfermion loop corrections to the chargino and
neutralino two--point functions, plus the counterterms from (s)fermion loops
needed to render them finite. Note that the sum of these explicit two--point
function corrections and counterterms has to be finite, since these are the
only one--loop corrections to $\tilde \chi f \tilde f$ vertices involving
(s)fermions $f', \tilde f'$ of a {\em different} $SU(2)$ multiplet. As usual,
the finiteness of the results is an important check on the calculation.

\setcounter{footnote}{0}

In \cite{Guasch:2002ez} all these contributions were absorbed in
corrections to the matrices $N, \, U$ and $V$:
\begin{equation}
\label{eq:dUVN1}
      \tilde{U} = U + \Delta U \, , \ \ 
      \tilde{V} = V + \Delta V \, ,  \ \
      \tilde{N} = N + \Delta N \, .
\end{equation}
Note that $\Delta U, \, \Delta V, \, \Delta N$ are {\em not} counter
terms; rather, they describe {\em finite} corrections to the {\em
  couplings} of charginos and neutralinos to sfermions and fermions,
  and to Higgs bosons.\footnote{A perhaps clearer, but bulkier, notation
  would be $\Delta (gU), \, \Delta (gV),\, \Delta(gN)$, which can be
  obtained from eqs.(\ref{eq:dUVN2}) by multiplying both sides with $g$;
  note that even the contributions to the chargino and neutralino
  couplings (\ref{neutcoup}) and (\ref{charcoup}) that originate from
  superpotential couplings have a factor of $g$ in front. For easier
  comparison we here stick to the notation introduced by Guasch et
  al. \cite{Guasch:2002ez}.} In our renormalization scheme, as in
ref.\cite{Guasch:2002ez}, the mixing matrices are not renormalized at
all, since we allow off--diagonal wave function renormalization
counter terms.

Explicitly, the corrections defining the effective couplings are given
by 
\begin{eqnarray} \label{eq:dUVN2} 
\Delta U_{i1} & = & U_{i1}\left(\frac{\delta g}{g} + \frac{1}{2} \delta Z^{+R}_{i}
\right)+U_{j1} \frac{1}{2} \delta Z^{+R}_{ji}\,,
\nonumber\\ 
\Delta U_{i2} & = & U_{i2} \left( \frac{\delta g}{g} + \frac{1}{2}\delta
Z^{+R}_{i} - \frac{1}{2} \frac{\delta M_{W}^{2}}{M_{W}^{2}} - \frac{\delta
  \cos \beta}{\cos \beta}  \right) + U_{j2}\frac{1}{2} \delta Z^{+R}_{ji}\,,
\nonumber\\  
\Delta V_{i1} & = & V_{i1} \left( \frac{\delta g}{g} + \frac{1}{2}\delta
Z^{+L}_{i} \right) + V_{j1} \frac{1}{2}\delta Z^{+L}_{ji}\,, 
\nonumber\\
\Delta V_{i2} & = & V_{i2}\left(\frac{\delta g}{g}+\frac{1}{2}\delta Z^{+L}_{i}-
\frac{1}{2}\frac{\delta {M_W}^2}{{M_W}^2} -\frac{\delta \sin
  \beta}{\sin\beta}\right) +V_{j2} \frac{1}{2} \delta Z^{+L}_{ji}\,,
\nonumber\\  
\Delta N_{\alpha 1} & = &  N_{\alpha 1}\left(\frac{\delta g}{g} 
    + \frac{1}{2}\delta Z^{R}_{\alpha} + \frac{\delta \tan \theta_W}{\tan
      \theta_W} \right) + \sum_{\beta\neq\alpha} {N_{\beta 1}
      \frac{1}{2}\delta Z^{R}_{\beta\alpha}} \,, 
\nonumber\\ 
\Delta N_{\alpha 2} & = &  N_{\alpha 2}\left(\frac{\delta g}{g}+ 
    \frac{1}{2}\delta Z^{R}_{\alpha}\right) +\sum_{\beta\neq\alpha}{N_{\beta 2}
   \frac{1}{2} \delta Z^{R}_{\beta\alpha}}\,,
\nonumber\\ 
\Delta N_{\alpha 3} & = &  N_{\alpha 3} \left( \frac{\delta g}{g} +
  \frac{1}{2}\delta Z^{R}_{\alpha} + \frac{1}{2} \frac {\delta M_W^2}
  {M_W^2} - \frac {\delta \cos\beta} {\cos\beta} \right) 
  + \sum_{\beta\neq\alpha} {N_{\beta 3} \frac{1}{2} \delta Z^{R}_{\beta\alpha}} \,,
\nonumber\\
\Delta N_{\alpha 4} & = &  N_{\alpha 4}\left(\frac{\delta g}{g} + 
    \frac{1}{2}\delta Z^{R}_{\alpha}+\frac{1}{2}\frac{\delta M_W^2}{M_W^2}-
    \frac{\delta \sin\beta}{\sin\beta}\right) + \sum_{\beta\neq\alpha}
    {N_{\beta 4}\frac{1}{2}\delta Z^{R}_{\beta\alpha}} \,.
\end{eqnarray}
In the first four equations, for $\Delta U$ and $\Delta V$, the subscript $j$
in the last term must always be different from $i$ appearing on the left--hand
sides, i.e. these are contributions from purely off--diagonal two--point
functions, just like the corresponding contributions to the $\Delta N$.  The
counterterms $\delta g$ and $\delta \tan \theta_W$ are given in Appendix A,
while $\delta \sin\beta$ and $\delta \cos\beta$ are given in Appendix
C. Finally, the diagonal and off--diagonal wave function counterterms, $\delta
Z_i$ and $\delta Z_{ij}$, are given at the end of Appendix D; here the
superscript $L$ and $R$ refer to chirality, while the superscript $+$
refers to charginos.

We emphasize that $\Delta N$, $\Delta U$ and $\Delta V$, as described above,
are process independent in the sense that they do not depend on the details of
the external state [the flavor of the (s)fermion or the identity of the Higgs
  boson].

An important feature of these corrections is that heavy sparticles do {\em
  not} decouple; to the contrary, the size of the corrections {\em increases}
logarithmically if some sfermions are much heavier than the charginos or
neutralinos whose couplings are being computed. This observation is especially
relevant in the light of recent LHC data which imply stringent lower bounds on
the masses of first and second generation squarks, but allow relatively light
superparticles with only electroweak interactions. As an illustration let us
first review the origin of these non--decoupling effects only including gauge
interactions.

Unbroken supersymmetry implies that the fermion--fermion--gauge boson
(gauge) coupling $g$ is the same as the corresponding
sfermion--fermion--gaugino coupling $\tilde{g}$. Since supersymmetry
is broken at a high scale (compared to the light fermion masses), the
heavier sfermions {\em and fermions} will decouple from the running of
$\tilde{g}$ below the sfermion mass scale $m_{\tilde f}$. However, the
corresponding fermion will still contribute to the running of
$g$. Therefore, $g$ and $\tilde{g}$ ``run apart'' below the highest
supersymmetry breaking scale. 

We use the running coupling $\alpha(M_Z)$ as tree--level input coupling, which
is defined at the fixed scale $M_Z$. Consider a spectrum where sleptons,
charginos and neutralinos have similar masses to each other, while squarks
have much higher masses, $m_{\tilde \ell} \sim m_{\tilde \chi} \ll m_{\tilde
  q}$. We are interested in processes at the slepton or neutralino mass
scale. On the other hand, full supersymmetry is attained only at scale
$m_{\tilde q}$. Since between scales $m_{\tilde \chi}$ and $m_{\tilde q}$ the
leptonic sector is still supersymmetric, leptons and sleptons contribute in
the same way to the running of $g$ and $\tilde g$ in that range. Altogether we
thus have:
\begin{eqnarray} \label{running}
\tilde g(m_{\tilde \chi}) &=& \tilde g(m_{\tilde q}) - \beta_{\ell,\tilde \ell} \log
\frac {m_{\tilde q}} {m_{\tilde \chi}}
\nonumber \\
& = & g(m_{\tilde q}) - \beta_{\ell,\tilde \ell} \log \frac {m_{\tilde q}} 
{m_{\tilde \chi}}
\nonumber \\
& = & g(M_Z) + \beta_{\ell,q} \log \frac {m_{\tilde \chi}} {M_Z} + 
\beta_{\ell,q,\tilde \ell} \log \frac {m_{\tilde q}} {m_{\tilde \chi}} - 
\beta_{\ell,\tilde \ell} \log \frac {m_{\tilde q}} {m_{\tilde \chi}}
\nonumber \\
& = & g(M_Z) + \beta_{\ell,q} \log \frac {m_{\tilde \chi}} {M_Z} + \beta_q \log
\frac {m_{\tilde q}} {m_{\tilde \chi}}\,.
\end{eqnarray}
Here the particles contributing to the gauge beta function have been listed
explicity as subscripts; in the last step we have used the fact that to
one--loop order the contributions of different (s)particles simply add up,
i.e. $\beta_{\ell, q, \tilde \ell} = \beta_{\ell, \tilde \ell} + \beta_q$.
Since SM fermions contribute with positive sign to the gauge beta functions,
both logarithmic factors in the last line of eq.(\ref{running}) are {\em
  positive} \cite{Cheng:1997sq, Katz:1998br}.

We use on--shell renormalization both in the electroweak and in the $\tilde
\chi$ sector. Our effective $\tilde \chi_i$ couplings can thus be understood
as running couplings at scale $m_{\tilde \chi_i}$. This is the appropriate
energy scale to describe $\tilde \chi_i$ decays, where the effective couplings
have been shown to reproduce the bulk of electroweak radiative corrections
\cite{dhx}, as well as for the annihilation of two neutralinos, which is the
main topic of the present investigation. It may not be the best choice for
sfermion decays, where these couplings have first been introduced
\cite{Guasch:2002ez}, nor for slepton production at a lepton collider via
$\tilde \chi$ exchange, where the $\tilde \chi$ is always quite far
off--shell.

The existence of Higgs superfields, whose superpotential couplings give rise
to matter fermion masses and whose fermionic members mix with the gauginos,
considerably complicates the picture. Fortunately the chirality structure
allows to distinguish ``gauge'' and ``superpotential'' contributions to the
$\tilde \chi f \tilde f$ couplings: the former couple $f_L$ to $\tilde f_L$
and -- for neutralinos coupling via hypercharge -- $f_R$ to $\tilde f_R$,
whereas the latter couple $f_L$ to $\tilde f_R$ and $f_R$ to $\tilde f_L$.

At one--loop level true Yukawa couplings $\lambda$ do not renormalize true
gauge couplings $g$. The above discussion of the ${\cal O}(g^3)$ corrections
to the ``gauge'' contribution of $\tilde \chi f \tilde f$ couplings therefore
indicates that there are {\em no} ${\cal O}(g \lambda^2)$ corrections to these
couplings. This is in fact true in the absence of gaugino--higgsino mixing,
or, to good approximation, for gaugino--like states, where ${\cal
  O}(g \lambda^2)$ corrections are suppressed by two factors of
higgsino--gaugino mixing angles.

However, the ``gauge'' couplings of higgsino--like states do receive
corrections ${\cal O}(g \lambda^2)$, where (s)top, (s)bottom and (s)tau
loops all can contribute. These corrections are suppressed by one
factor of a higgsino--gaugino mixing angle $\epsilon$; however, since
this is true already for this tree--level coupling, the relative size
of the correction is not suppressed. Technically, these corrections
involve the ``vector'' parts of diagonal as well as off--diagonal
$\tilde \chi$ two point functions, as well as the ``scalar'' parts of
off--diagonal two--point functions; see eq.(\ref{Sigma}) for the
decomposition of fermionic two--point functions. In the latter case,
the product $g \lambda_f m_f$ has been counted as ${\cal
  O}(\lambda_f^2 \epsilon)$, where the factor $\epsilon$ describing
suppression by higgsino--gaugino mixing arises from the combination of
the factor $M_W$ in $m_f \propto \lambda_f M_W / g$ and the $m_{\tilde
  \chi}$ factors in the expressions for the off--diagonal wave
function counterterms given in eqs.(\ref{eqzl}) and (\ref{eqzr}).

Since there are no unsuppressed ${\cal O}(g \lambda^2)$ corrections to
the ``gauge'' couplings of gaugino--like states, these corrections
cannot be understood in terms of running $\tilde g$
couplings. Instead, in the framework of the effective theory defined
by integrating out heavy sfermions, they can be interpreted as
contributions to the running of those (off--diagonal) elements in the
chargino and neutralino mass matrices that mix higgsino-- and
gaugino--like states.

This interpretation is supported by the observation that the
``superpotential'' couplings $\tilde \lambda$ of gaugino--like states
receive one-loop corrections ${\cal O}(\lambda g^2), \, {\cal
  O}(\lambda^3)$, where again all heavy (s)fermions contribute, that
are only linearly suppressed by gaugino--higgsino mixing. Since these
tree--level couplings suffer a similar suppression, the relative size
of the corrections is again not suppressed. In contrast, the
``superpotential'' coupling of higgsino--like states only receives
unsuppressed ${\cal O}(\lambda^3)$ corrections, while ${\cal
  O}(\lambda g^2)$ corrections are suppressed by {\em two} powers of
higgsino--gaugino mixing angles; moreover, the for our application
most relevant coupling $\propto \tilde \lambda_\tau$ in this case does
not receive unsuppressed corrections from the (s)top sector. This is
consistent with the observation that only true Yukawa couplings
contribute to the beta functions of other Yukawa couplings, if one
restricts oneself to matter (s)fermion loops; since matter loops
contribute with positive sign to the beta functions of superpotential
couplings, the corresponding corrections are positive, just as in the
pure gauge sector.\footnote{In the full theory there are ${\cal O}(g^2
  \lambda)$ corrections to the beta function of $\lambda$, but these
  are due to diagrams where members of gauge superfields run in the
  loop. These contributions cannot be included consistently into our
  effective couplings.}

Finally, the ``gauge'' couplings of higgsino--like states also receive
${\cal O}(g^3)$ corrections, again suppressed by one factor of
higgsino--gaugino mixing; these originate both from the running of the
``gauge'' coupling $\tilde g$ and from ${\cal O}(g^2)$ corrections to
gaugino--higgsino mixing.  Similarly, the ``superpotential'' couplings
$\tilde \lambda$ of gaugino--like states receive ${\cal O}(\lambda^3)$
corrections, suppressed by one factor of higgsino--gaugino mixing,
that can be understood from the running of $\tilde \lambda$.

Altogether, the tree--level and logarithmically enhanced $\tilde \chi \ell
\tilde \ell$ couplings, where $\ell$ stands for a lepton, can be summarized by
table~\ref{tab:coup}, where we have not distinguished between $g$ and $\tilde
g$, nor between the superpotential coupling $\lambda$ and the corresponding
contribution $\tilde \lambda$ to $\tilde \chi f \tilde f$ couplings. Of
course, our effective couplings also include terms that are not enhanced
logarithmically for large sfermion mass. The corrections that vanish as some
inverse power of the mass of the sfermion in the loop do not follow the
pattern of this table; for example, there are ${\cal O}(g \lambda_t^2
m^2_{\tilde \chi} / m^2_{\tilde t})$ corrections to the ``gauge'' couplings of
gaugino--like $\tilde \chi$ states. However, these decoupling corrections are
usually small in our numerical applications.

\begin{table}[h!]
\begin{center}
\begin{tabular}{|c||c|c||c|c|}
\hline
& \multicolumn{2}{|c||}{$\tilde \chi \simeq$ higgsino} &
\multicolumn{2}{|c|}{$\tilde \chi \simeq$ gaugino} \\
coupling & tree--level & one--loop & tree--level & one--loop \\
\hline 
$\tilde \chi \ell_L \tilde \ell_L, \ \tilde \chi \ell_R \tilde \ell_R$ &
${\cal O}(\epsilon g)$ & 
${\cal O}(\epsilon g^3, \ \epsilon g \lambda^2)$ & ${\cal O}(g)$ & $ {\cal
  O}(g^3, \ \epsilon^2 g \lambda^2)$ \\
$\tilde \chi \ell_L \tilde \ell_R, \ \tilde \chi \ell_R \tilde \ell_L$ &
${\cal O}(\lambda_\ell)$ & 
${\cal O}(\lambda_\ell^3, \ \lambda_\ell \lambda_b^2, \ \epsilon^2
\lambda_\ell g^2, \ \epsilon^2 \lambda_\ell \lambda_t^2)$ & $ {\cal O}(\epsilon
\lambda_\ell)$ & ${\cal O}(\epsilon \lambda_\ell g^2, \ \epsilon \lambda_\ell
\lambda^2)$ \\
\hline
\end{tabular}
\end{center}
\caption{Comparison of tree--level and one--loop contributions to
  slepton--lepton currents of different chiral structure, for higgsino-- and
  gaugino--like charginos and neutralinos $\tilde \chi$. Here $g$
  stands for an electroweak gauge coupling, $\lambda_\ell$ for the Yukawa
  coupling of lepton $\ell$, and $\epsilon$ for one factor of
  gaugino--higgsino mixing, which is assumed to be small here. A coupling
  $\lambda$ without subscript means that all superpotential couplings
  contribute without additional suppression. We do not distinguish
  between $g$ and $\tilde g$, nor between $\lambda$ and $\tilde
  \lambda$, in this table.}
 \label{tab:coup}
\end{table}

We repeat that the non--decoupling corrections from the gauge sector
are enhanced by multiplicity factors in addition to possibly large
logarithmic factors. Moreover, the effective couplings capture all
corrections $\propto \lambda_\tau \lambda_b^2, \lambda_\tau
\lambda_t^2$ to tree--level couplings $\propto \lambda_\tau$. They do,
however, {\em not} capture all corrections $\propto \lambda_t^3$ to
tree--level couplings $\propto \lambda_t$; since the latter also
receive QCD corrections, which are also not included in our effective
couplings, their application to $\tilde \chi q \tilde q$ couplings
involving superpotential couplings is probably of limited usefulness.

\setcounter{footnote}{0}

As mentioned above, the corrections described by eqs.(\ref{eq:dUVN2}) can all
be computed from two--point functions; see the Appendices for details. At a
similar level of effort one can include corrections to the (pole) masses of
the charginos and neutralinos. Of course, these are always computed from
two--point functions only; we can, and do, therefore include the {\em
  complete} correction to the masses, including corrections from the gauge and
Higgs sectors in addition to those from the matter (s)fermion
sector.\footnote{Computing the corrections to neutralino and chargino masses
  entails again computation of the counterterms $\delta M_W,\, \delta g, \,
  \delta \cos \theta_W$ and $\delta \cos\beta$, now including {\em all}
  one--loop corrections. In the calculations of the one--loop corrected masses
  these counterterms will therefore have different numerical values than those
  appearing in eqs.(\ref{eq:dUVN2}), where {\em only} corrections from
  (s)fermion loops must be included in order to ensure finiteness of the
  effective couplings.} We do so using a recently introduced \cite{CDKX}
variant of the on--shell scheme, where we consider the masses of the (more)
wino--like chargino and of the the (most) bino-- and higgsino--like
neutralinos as inputs, which are not altered by the corrections. In
ref.\cite{CDKX} it was shown that the counterterms to $M_1$, $M_2$, and $\mu$
then remain well behaved in most of the parameter space, in contrast other
versions of the on--shell scheme \cite{Fritzsche:2002bi}. As a result $\Delta
N$, $\Delta U$ and $\Delta V$ also remain well behaved even when $M_2 =
|\mu|$.

The expressions (\ref{eq:dUVN2}) were originally introduced
\cite{Guasch:2002ez} as corrections to couplings of charginos and neutralinos
to fermions and sfermions. However, in addition, the corrections $\Delta
N_{\alpha 1}$, $\Delta N_{\alpha 2}$, $\Delta U_{i1}$ and $\Delta V_{i1}$ can
be used to (partly) absorb radiative corrections to all the vertices which
involve a gaugino component of a chargino or neutralino but do {\em not}
include a gauge boson. These include couplings of neutralinos and charginos to
Higgs bosons, the {\em gaugino} component of which we also modify. However,
the corrections $\Delta N_{\alpha 3}, \, \Delta N_{\alpha 4}, \, \Delta
U_{i2}, \, \Delta V_{i2}$ should {\em not} be applied to these Higgs vertices,
but {\em only} to vertices involving a (massive) fermion and one of its
superpartners; as discussed above, part of the latter corrections can be
understood as process independent corrections to chargino and neutralino
couplings originating from the superpotential, whereas the couplings of Higgs
bosons to neutralinos and charginos are pure gauge couplings in the limit of
exact supersymmetry.\footnote{Another way to see this is to note that it makes
  little sense to include higgsino two--point function corrections to a
  Higgs--higgsino--gaugino vertex, without including closely related
  two--point function corrections on the Higgs boson line.} Similarly, a
vertex with a gauge boson involves a true gauge coupling. The running of the
corresponding gauge coupling (the $U(1)_Y$ and $SU_{L}(2)$ gauge couplings in
our context) takes care of all process independent radiative corrections to
these vertices, which should therefore be written in terms of the {\em
  original} $U, \, V, \, N$ matrices, {\em without} the corrections of
eqs.(\ref{eq:dUVN2}).

\begin{figure}[!ht]
\epsfig{file=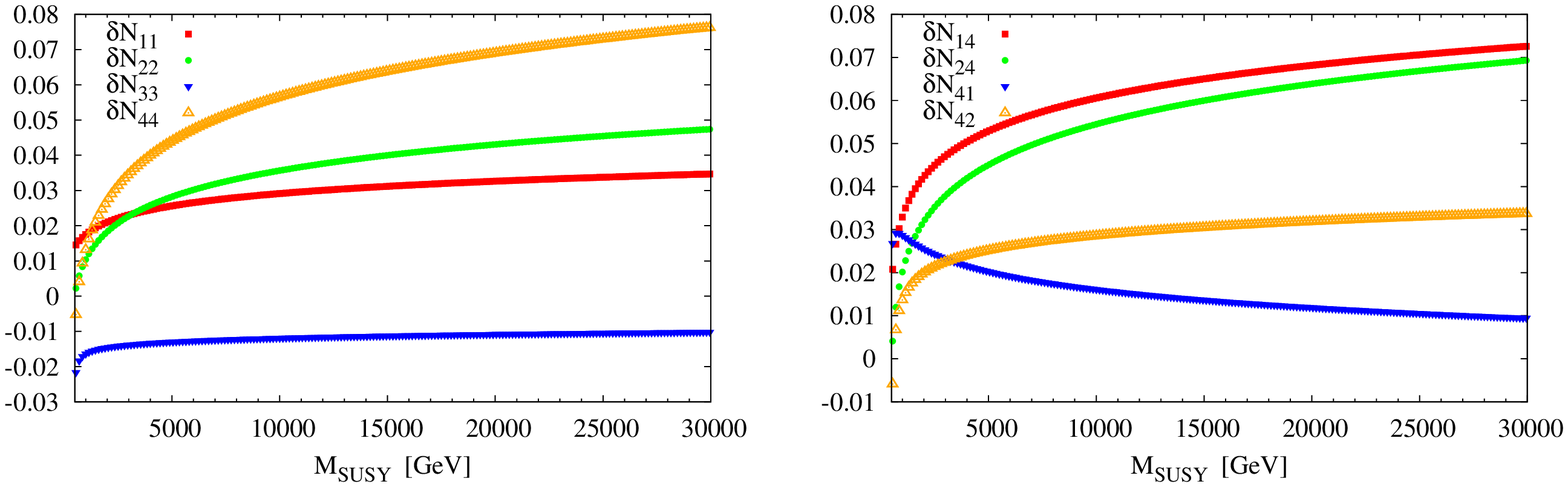, width=\linewidth, height=6 cm} 
$~~~~~~~~~~~~~~~~~~~~~~~~~~~~~~ \text{(a)}  \hspace{7.75cm} \text{(b)}$
\caption{$\delta N_{\alpha\beta } = \dfrac{\Delta N_{\alpha\beta}}
  {N_{\alpha\beta}}$, as given by eq.(\ref{eq:dUVN2}) against $M_{\rm
    SUSY}$, where $M_{\rm SUSY}$ denotes the common soft SUSY breaking
  sfermion mass.}
\label{fig:dns}
\end{figure}

Before presenting numerical results for corrections to neutralino annihilation
cross sections, we wish to show some results for the corrections to neutralino
couplings, i.e. the $\Delta N_{\alpha\beta}$. In Fig.~\ref{fig:dns} we plot
the {\em relative} size of these corrections, $\Delta N_{\alpha\beta} /
N_{\alpha\beta}$, for selected values of $\alpha$ and $\beta$, against a
common (i.e. generation and flavor independent) soft SUSY breaking sfermion
mass $M_{\rm SUSY}$. We assume the following parameters (all mass
parameters are in GeV): 
\beq
\tan \beta = 10,~ M_A = 500,~ M_1 = 100,\\ 
M_2 = 300,~ M_3 = 1200,~ \mu=600, A=0. 
\eeq
Since $\mu - M_{1,2} \gg M_Z$, gaugino--higgsino mixing is suppressed,
so the results of table~\ref{tab:coup} are applicable. Specifically,
$\tilde \chi_1^0$ is bino--like, $\tilde \chi_2^0$ and $\tilde
\chi_1^+$ are wino--like, and $\tilde \chi_3^0, \, \tilde \chi_4^0$
and $\tilde \chi_2^+$ are higgsino--like.

The left frame shows corrections to couplings that are not suppressed
by gaugino--higgsino mixing. Recall that the subscript $\alpha$ in
$N_{\alpha \beta}$ refers to the mass eigenstate (ordered by
increasing mass), while $\beta = 1,\, 2,\, 3,\, 4$ refers to the bino,
wino, down--higgsino and up--higgsino component, respectively.
Clearly the biggest correction are those to $N_{44}$, which involve
superpotential couplings of the ``up--type'' higgsino, in particular
the top coupling $\lambda_t$.  There are unsuppressed ${\cal
  O}(\lambda_t^2)$ contributions to $\delta N_{44}$; this case is
analogous to the ${\cal O} (\lambda_\ell^3)$ corrections to the
$\tilde \chi \ell_L \tilde \ell_R$ couplings in
table~\ref{tab:coup}. For the given, rather small, value of
$\tan\beta$, $\delta N_{33}$ is far smaller, since it only receives
contributions $\propto \lambda_b^2, \, \lambda_\tau^2$, plus ${\cal
  O}(\epsilon^2)$ contributions. Not surprisingly, if $\tan\beta \sim
m_t/m_b$, $\delta N_{33}$ is of similar size as $\delta N_{44}$.

Among the corrections to the dominant couplings of the gaugino--like
neutralinos, $\delta N_{22}$ increases significantly faster with
increasing $M_{\rm SUSY}$ than $\delta N_{11}$ does simply because the
$SU(2)$ gauge coupling is larger than the $U(1)_Y$ coupling; this is
only partly compensated by the fact that $SU(2)$ singlet (s)fermions
do contribute to $\delta N_{11}$.

The right frame of Fig.~\ref{fig:dns} shows corrections to couplings that are
${\cal O}(\epsilon)$ at tree--level. The biggest corrections are to $N_{14}$
and $N_{24}$, involving ``superpotential'' couplings of gaugino--like
states. These corrections are dominated by ${\cal O}(\lambda_t^2)$ running
coupling effects, and are in fact of similar magnitude as $\delta N_{44}$
shown in the left frame; recall that we are showing the {\em relative} size of
the one--loop correction here, so that the ${\cal O}(\epsilon)$ factor
appearing in all couplings shown in the right frame cancel out. 

The corrections to the ``gauge'' couplings of the higgsino--like states
$N_{41}$ and $N_{42}$ are much smaller. These receive running coupling
contributions proportional to the squared $U(1)_Y$ and $SU(2)$ gauge coupling,
respectively, but are much smaller than $\delta N_{11}$ and $\delta N_{22}$
shown in the left frame. This indicates that the corrections to
gaugino--higgsino mixing, which are dominated by ${\cal O}(\lambda_t^2)$
contributions, tend to reduce the couplings in this case. In fact, the total
coefficient of the correction $\propto \log(M_{\rm SUSY})$ to $\delta
N_{41}$ is negative. The fact that the overall correction nevertheless remains
positive even at $M_{\rm SUSY} = 30$ TeV illustrates that the terms that
approach constants for large sfermion masses can also be very important.

This completes our discussion of the effective couplings. Before
presenting numerical results for the annihilation cross section and
relic density of the lightest neutralino, we briefly describe the
implementation of the effective couplings using publicly available
software.

\section{Implementation}
\label{sec:Imp}
\setcounter{footnote}{0}

\begin{figure}[!ht]
\epsfig{file=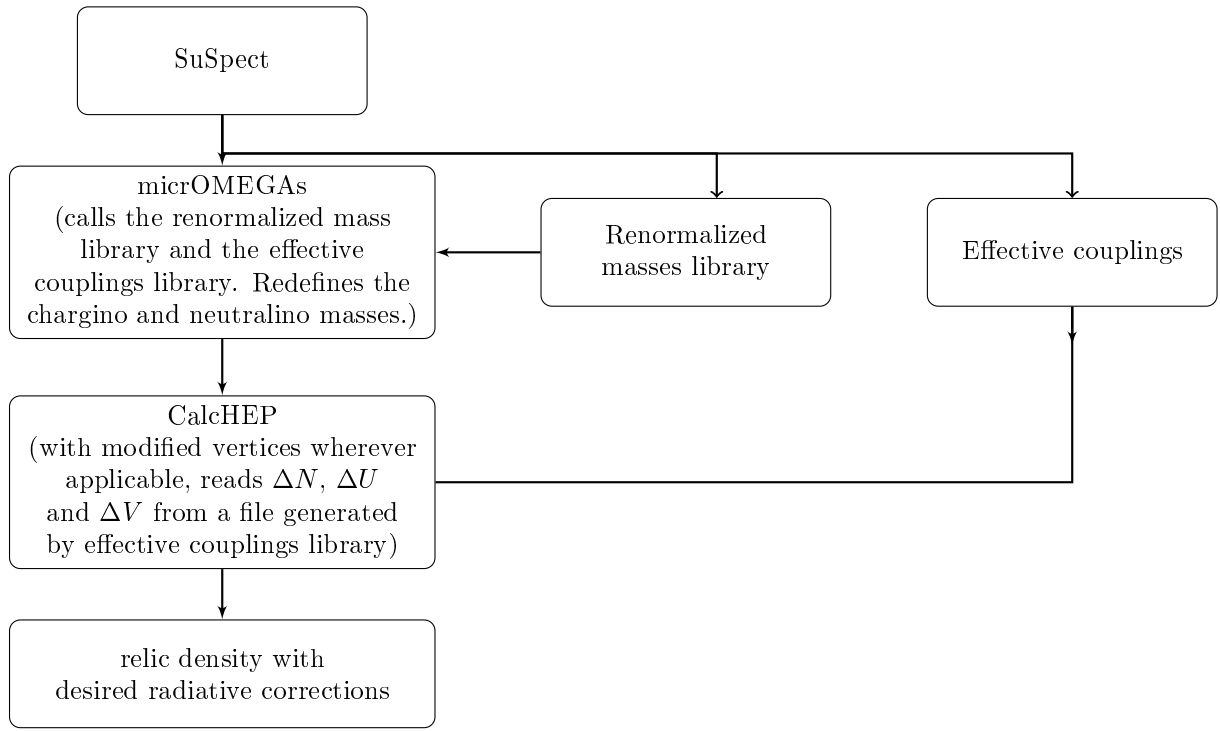, width=\linewidth} 
\caption{Flowchart describing the interaction of the publicly
  available codes \texttt{SuSpect} and \text{micrOMEGAs} with our
  subroutines calculating corrections to $\tilde \chi$ masses and
  couplings.}
\label{fig:chart}
\end{figure}

In order to predict the relic density of the lightest neutralino at one--loop
level, we have modified one of the publicly available codes for relic density
calculation, \texttt{micrOMEGAs} (version 2.2.CPC.i)
\cite{Belanger:2008sj}. The algorithm describing the flow of relevant
parameters is sketched in Fig.~\ref{fig:chart}.

We use \texttt{SuSpect} \cite{Djouadi:2002ze} to compute the ``tree--level''
spectrum of superparticles and Higgs bosons. \texttt{SuSpect} actually
includes various loop corrections in converting input $\overline{\rm DR}$
parameters into on--shell parameters and, hence, physical masses. However, in
the neutralino--chargino sector these corrections are complete only in the
absence of higgsino--gaugino mixing. Since we implement full one--loop
corrections to all $\tilde \chi$ masses in the on--shell scheme, we had to
slightly modify \cite{CDKX} the $\tilde \chi$ mass matrices used in
\texttt{SuSpect}, using on--shell $W$ and $Z$ masses and the on--shell
definition of the weak mixing angle $\theta_W$.

The output generated by \texttt{SuSpect} (in the format of the
Supersymmetric Les Houches Accord, SLHA \cite{SLHA}) is used as input
of \texttt{micrOMEGAs}. This program first computes two external
libraries, written by us, which calculate the effective couplings and
the renormalized masses, respectively. We have used \texttt{FeynArts,
  FormCalc} \cite{Hahn:2000kx,Hahn:2001rv, Hahn:2006zy} and
\texttt{LoopTools} \cite{Hahn:1998yk} to compute the relevant diagrams
and counterterms. Feynman gauge has been used in the calculation of
one--loop corrections to $\tilde \chi$ masses; the effective couplings
do not include loops involving gauge bosons. For regularization we
have used the constrained differential renormalization method
\cite{delAguila:1998nd}. At one--loop level this method has been
proven to be equivalent \cite{Hahn:1998yk} to regularization by the
dimensional reduction method \cite{DR}. Note that, unlike the
effective couplings, the renormalized masses involve contribution from
all MSSM (s)particles; the two external libraries therefore use
independent calculations of the relevant counterterms, as further
described in the Appendices. In the loop diagrams, we for simplicity
used running $\overline{\rm DR}$ sfermion masses defined at the
electroweak scale, as given by \texttt{SuSpect};\footnote{The reason
  is that the version of \texttt{SuSpect} we were using implemented
  certain ${\cal O}(\lambda_t^2)$ corrections to stop masses, but not
  to sbottom masses, thereby creating spurious hard breaking of
  $SU(2)$ in the stop--sbottom sector, with sizable effects on certain
  electroweak counterterms.} the use of a different renormalization
scheme for the sfermion masses inside the loops affects our results
only at two--loop level. Moreover, we ignore $L-R$ mixing for first
and second generation sfermions, and do not include any flavor mixing.

\texttt{MicrOMEGAs} calculates the $\tilde \chi_1^0$
(co--)annihilation cross sections with the help of \texttt{CalcHEP}
\cite{calchep}. We modified \texttt{CalcHEP} to read the effective
couplings $\Delta U, \, \Delta V$ and $\Delta N$. These are used in
{\em all} $\tilde \chi f \tilde f$ couplings; the corrections to
gaugino couplings $\Delta U_{i1}, \, \Delta V_{i1}, \, \Delta
N_{\alpha 1}$ and $\Delta N_{\alpha 2}$ are also used in $H \tilde
\chi \tilde \chi$ vertices, but the corrections to higgsino couplings
$\Delta U_{i2}, \, \Delta V_{i2}, \, \Delta N_{\alpha 3}$ and $\Delta
N_{\alpha 4}$ are {\em not} used in these vertices, and the couplings
between $\tilde \chi$ states and gauge bosons are not modified at all.

We already mentioned that we use on--shell renormalization in the
electroweak sector. This refers to the $W$ and $Z$ masses, the weak
mixing angle $\theta_W$ as well as the running electromagnetic
coupling $\alpha(M_Z)$. For consistency we thus have to use on--shell
values for these quantities also in the tree--level calculation. This
required additional changes of \texttt{micrOMEGAs}, which uses an
on--shell $M_Z$, but the $\overline{\rm MS}$ value of $\alpha(M_Z)$,
which is nearly 1\% larger than the on--shell value, and an effective
weak mixing angle such that $\sin^2 \theta_W \simeq 0.231$ (compared
to $0.222$ in the on--shell scheme). We use this modified version of
\texttt{micrOMEGAs} {\em only} for the one--loop corrected cross
section. Since \texttt{micrOMEGAs} currently is the ``industry
standard'' for calculations of the $\tilde \chi_1^0$ relic density, we
use its original version when evaluating the final corrections to the
(co--)annihilation cross sections and relic density. The relative
corrections to the annihilation cross sections are thus defined as
\beq \label{delsig}
\delta \sigma = \frac { \sigma^{\rm tree, \ on-shell} + \Delta
  \sigma^{\rm 1-loop} - \sigma^{\rm tree, \ orig} } {\sigma^{\rm tree,\
    orig} }\,, 
\eeq
where $\sigma^{\rm tree, \ orig}$ is the prediction of the original
version of \texttt{micrOMEGAs} using $\overline{\rm MS}$ values of
$\alpha(M_Z)$ and $\theta_W$.\footnote{Recall, however, that we
  slightly changed the off--diagonal entries in the chargino and
  neutralino mass matrices in \texttt{SuSpect}; the resulting $\tilde
  \chi$ masses and mixing angles are also used by the ``original''
  \texttt{micrOMEGAs} in our calculation.} We emphasize that using
$\sigma^{\rm tree, \ on-shell}$, computed with on--shell values for
$\theta_W$ and $\alpha(M_Z)$, as reference value would have led to
significantly {\em larger} loop corrections $\delta \sigma$ for
bino--like $\lneut$, largely because $\alpha(M_Z)$ is smaller in the
on--shell scheme, as mentioned above. Note finally that we use
on--shell values of $\alpha(M_Z)$ and $\theta_W$ {\em only} in those
vertices which we correct by the effective couplings. In all other
vertices we continue to use \texttt{micrOMEGAs} default values of
parameters; this includes all couplings of gauge bosons, as well as
couplings of Higgs bosons to SM particles. This ensures that $\delta
\sigma$ defined in eq.(\ref{delsig}) only shows effects of loop
corrections we include but standard \texttt{micrOMEGAs} doesn't
include.\footnote{For example, changing $\alpha(M_Z)$ and $\theta_W$
  everywhere would also lead to changes of the Higgs spectrum, which
  can lead to large changes of the annihilation cross section near
  Higgs poles; these would be entirely due to changes of input
  parameters, not due to radiative corrections.}

\section{Numerical Results}
\label{sec:Num}

Before presenting numerical results for a benchmark scenario, a qualitative
discussion of the usefulness of the effective couplings for the calculation of
the $\tilde \chi_1^0$ relic density might be in order.

We expect the effective couplings to be most useful for dominantly bino--like
neutralino \cite{Boudjema:2011ig}. A bino--like neutralino dominantly
annihilates through $t$ and $u$ channel sfermion exchange to a pair of
(lighter) fermions. Effective couplings absorb significant radiative
corrections in both vertices. These will likely dominate the total electroweak
corrections whenever squarks are significantly heavier than neutralinos and
sleptons, since this leads to logarithmic enhancement of our corrections, as
discussed above. The most important final states will then be $\ell^+ \ell^-$
pairs. Since the cross section is $\propto |N_{11}|^4$, the relative size of
the correction can be estimated as $4 \delta N_{11}$.

However, given existing lower bounds on slepton masses \cite{PDG},
pure slepton exchange usually gives too small an annihilation cross
section, i.e. too large a $\tilde \chi_1^0$ relic density (in standard
cosmology). This can be corrected through co--annihilation with a
slepton $\tilde \ell$, which also involves the $\tilde \chi_1^0 \ell
\tilde \ell$ vertex.

The annihilation cross section of bino--like neutralinos can also be
enhanced sufficiently (or even by too much) if there is a neutral
Higgs boson with mass near twice the neutralino mass, the CP--odd
state $A$ being most efficient for this purpose. We also use an
effective coupling for the gaugino components in the $\tilde \chi
\tilde \chi$~Higgs vertices. The flavor independent contributions to
the two point function corrections to the gaugino components remain
finite, and can absorb significant corrections to these processes.

Given lower bounds on chargino masses, higgsino-- or wino--like
neutralinos will annihilate dominantly into gauge boson pairs $W^+
W^-, \, Z Z$.  These cross sections involve true gauge interactions,
i.e. logarithmically enhanced loop corrections can be absorbed into
the running of the $SU(2)$ [and, in case of higgsinos, $U(1)_Y$] gauge
coupling. Our effective couplings can still capture leading
corrections to subleading $f \bar f$ final states for the annihilation
of wino--like neutralinos; they are also likely to be quite important
in the vicinity of possible $s-$channel Higgs resonances.

Let us now turn to a discussion of numerical results, based on the following
benchmark scenario:
\begin{eqnarray} \label{bench}
m_{\tilde q} &=& 1.5 \ {\rm TeV}, M_3 = 1.2 \ {\rm TeV}, m_{\tilde l_L} = 0.55
\ {\rm TeV}, m_{\tilde l_R} = 0.50 \ {\rm TeV}, \nonumber \\
\mu &=& 0.6 \ {\rm TeV}, \ m_A = 0.5 \ {\rm TeV}, \ A_3 = 1.0 \ {\rm TeV},
\ \tan(\beta)(M_Z) = 10\,.
\end{eqnarray}
All parameters except the on--shell mass $m_A$ of the CP--odd Higgs boson are
running $\overline{\rm DR}$ parameters, in most cases defined at the scale of
electroweak symmetry breaking (EWSB), which in SuSpect is defined as
$\sqrt{m_{\tilde t_1} m_{\tilde t_2}} \simeq 1.5$ TeV in our benchmark
scenario. This choice minimizes one--loop corrections to the effective
potential for the neutral Higgs fields \cite{Gamberini:1989jw,
  deCarlos:1993yy}. Different values will be chosen for the $SU(2)$ and
$U(1)_Y$ gaugino masses $M_2$ and $M_1$ in subsequent figures, as stated
below.

The input value of $\tan \beta$ is defined at scale $M_Z$, which is the
SuSpect default. We have used $\tan \beta ^{\overline {\rm DR }}$ at the EWSB
scale, approximately equal to 9.6, for our calculations. We do not include any
plots against $\tan \beta$, since neutralino masses and mixing matrices are
not very sensitive to it.

In all parameter scans we ensure that $\lneut$ is the lightest
supersymmetric particle. Since we assume exact $R$ parity, $\lneut$ remains
a dark matter candidate in these regions of parameter space. However, since
the main goal of this paper is to show the relative size of the radiative
corrections captured by effective couplings, we do not restrict ourselves to
regions of parameter space giving the correct relic density in standard
cosmology. Finally, in all figures showing $\lneut$ annihilation
cross--sections, these have been calculated assuming the relative velocity
between the annihilating neutralinos in their center of mass frame to be $v =
0.3$, which is typical for the epoch when the $\lneut$ decoupled from
SM particles.

\subsection{$M_1$ dependence}
\setcounter{footnote}{0}

\begin{figure}[!ht]
\epsfig{file=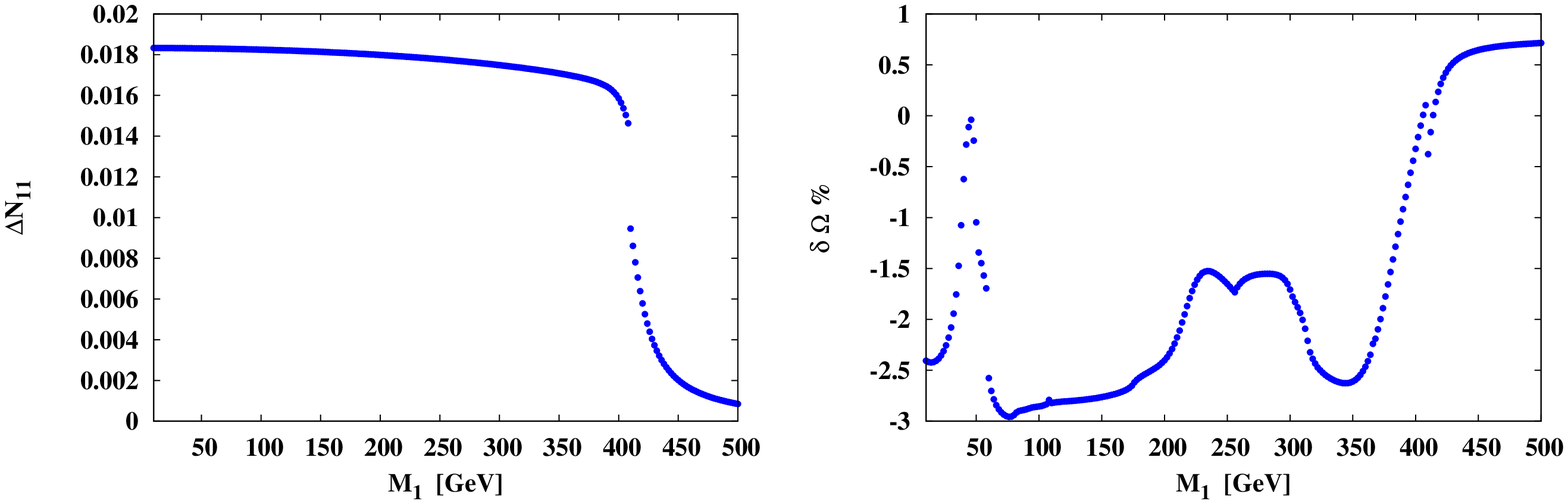, width=\linewidth, height=6 cm} \\
\hspace*{3cm} (a)  \hspace*{7.75cm} (b) \\
\epsfig{file=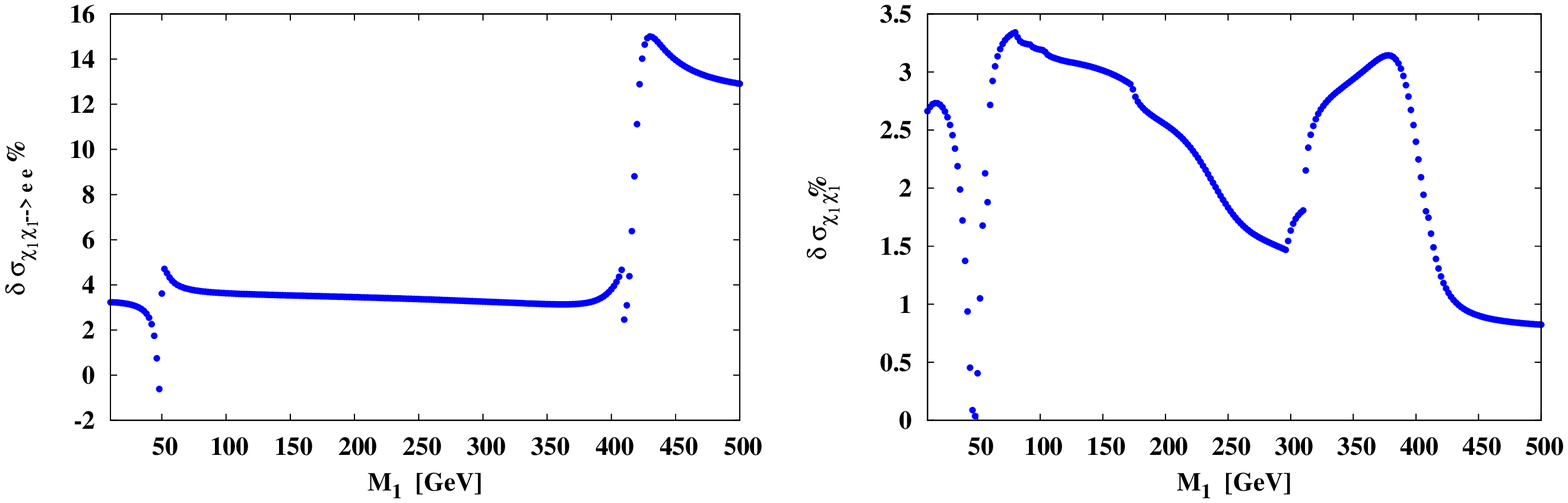, width=\linewidth, height=6 cm} 
$~~~~~~~~~~~~~~~~~~~~~~~~~~ \text{(c)}  \hspace{7.75cm} \text{(d)}$
\caption{Effect of radiative corrections as function of the bino mass
  parameter $M_1$ (in the $\overline{\rm DR}$ scheme at the
  electroweak symmetry breaking scale), for our benchmark point
  (\ref{bench}) with $M_2 = 0.4$ TeV. We show the effective coupling
  $N_{11}$ defined in eqs.(\ref{eq:dUVN2}) (top left); the relative
  correction to the predicted $\tilde \chi_1^0$ relic density (top
  right); the relative correction to $\sigma(\tilde \chi_1^0 \tilde
  \chi_1^0 \rightarrow e^+ e^-)$ (bottom left); and the relative
  correction to the total $\tilde \chi_1^0$ annihilation cross section
  (bottom right). The last two plots are for fixed relative velocity
  $v = 0.3$. All relative corrections have been computed as in
  eq.(\ref{delsig}).}
\label{fig:m1} 
\end{figure}

In Fig.~\ref{fig:m1}a we plot $\Delta N_{11}$, the correction to the
effective bino coupling of $\tilde \chi^{0}_{1}$, against $M_1$, for
$M_2 = 400$ GeV. Note that we show the total correction here, not the
relative correction shown in Fig.~\ref{fig:dns}. For $M_1 < 410$ GeV,
$\tilde \chi_1^0$ is bino--like and $\Delta N_{11} \simeq 0.018$ is
essentially independent of $M_1$. In this region of parameter space it
is dominated by RG--like ${\cal O}(g_Y^2)$ contributions, where $g_Y$
is the $U(1)_Y$ gauge coupling, and is thus positive. For $M_1 > 410$
GeV, $\tilde \chi_1^0$ becomes wino--like. This has little influence
on the relative size of the correction to the bino coupling, but,
since $|N_{11}|$ becomes much smaller than in the region of bino--like
$\tilde \chi_1^0$, $\Delta N_{11}$ is also greatly reduced.
Correspondingly, in this part of parameter space $\Delta N_{11}$ no
longer gives the most important correction to $\tilde \chi_1^0$
annihilation.

The relative size of the correction to the $\tilde \chi_1^0$ pair annihilation
cross section into an $e^+e^-$ pair, defined as in eq.(\ref{delsig}), is shown
in Fig.~\ref{fig:m1}c. Over most of parameter space this cross section
receives its dominant contributions from selectron (in particular, $\tilde
e_R$, due to the larger hypercharge) exchange in the $t-$ and
$u-$channels. For $M_1 \leq 400$ GeV, where $\tilde \chi_1^0$ is bino--like,
the correction is then essentially given by $4 \Delta N_{11}$; the relative
correction shown is smaller than this, because our reference cross section
does not use on--shell values for $\alpha(M_Z)$ and $\theta_W$, as explained
in eq.(\ref{delsig}). This indicates that it is indeed advantageous to use the
$\overline{\rm MS}$ value of $\alpha(M_Z)$ if one is performing a tree--level
calculation, since it reduces the size of the radiative
corrections.\footnote{The correction is further reduced because the small wino
  coupling of $\lneut$, whose magnitude is also increased by the correction
  $\Delta N_{12}$, reduces the overall $\tilde e_L e \lneut$
  coupling. However, this effect is numerically not very important, since the
  wino coupling remains very small, and the $\tilde e_L$ exchange contribution
  is about 20 times smaller than the $\tilde e_R$ exchange contribution; note
  that these contributions do not interfere in the limit $m_e \rightarrow 0$.}

An exception occurs for $M_1 \simeq 47$ GeV, where $\tilde \chi_1^0$
annihilation can occur through the exchange of an almost on--shell $Z$
boson. The very large enhancement factor $(M_Z / \Gamma_Z)^2 \simeq 1300$
overcompensates the coupling suppression $|N_{i3}|^{2}-|N_{i4}|^{2}$ of this
contribution to the matrix element. Since we do not modify the $Z \lneut
\lneut$ coupling, the size of the correction decreases at the $Z$ pole.

As noted above, for $M_1 \geq 410$ GeV $\lneut$ becomes
wino--like. Here the corrections, which are now dominated by terms
$\propto \Delta N_{12}$, become much larger; note that the cross
section is now dominated by $\tilde e_L$ exchange. We will see in
Fig.~\ref{fig:m2}a that $|\delta N_{12}| \simeq 1.3\%$ increases the
absolute size of the wino coupling, again due to RG--like
effects. Moreover, recall that our tree--level calculation uses
on--shell values of both $\alpha(M_Z)$ and $\theta_W$. As a result,
our tree--level $SU(2)$ gauge coupling is about $1.5\%$ larger than
that used by \texttt{micrOMEGAs}. Altogether our $\tilde e_L e \lneut$ wino
coupling is therefore about $3\%$ larger than in \texttt{micrOMEGAs}, leading
to an about $12\%$ enhancement of the $\lneut$ annihilation cross
section into $e^+e^-$ pairs in the deep wino--like region. Note that
for wino--like $\lneut$, our choice of tree--level couplings leads to
{\em smaller} radiative corrections than the choice made in
micrOMEGAs. This is in contrast to experience with LEP observables,
where the ``effective'' \cite{theta_eff} or $\overline{\rm MS}$
\cite{theta_MSbar} $\theta_W$\footnote{These two definitions of the
  weak mixing angle differ conceptually, but are numerically very
  similar.} leads to smaller radiative corrections than the
electroweak on--shell scheme. Recall, however, that for bino--like
$\lneut$ the use of $\overline{\rm MS}$ values for $\theta_W$ and
$\alpha(M_Z)$ reduces the size of radiative corrections.

The relative correction to the {\em total} $\lneut$ annihilation cross
section, summed over all kinematically accessible final states, is shown in
Fig.~\ref{fig:m1}d. It exhibits a considerably more complicated dependence on
$M_1$ than the cross section for the $e^+e^-$ final state does. The reason is
that different final states become important for different ranges of $M_1$.

Near the $Z$ pole the total annihilation cross section is again dominated by
$Z$ exchange, and the correction is small. For slightly larger $M_1$, the
$s-$channel exchange of the lighter CP--even Higgs state $h$ becomes (nearly)
resonant. As noted earlier, we correct the gaugino component of the $h \lneut
\lneut$ coupling. Since the $h$ exchange matrix element is linear in this
coupling, the correction is considerably smaller here than in regions of
parameter space where the total annihilation cross section is dominated by
$\tilde \ell$ exchange in the $t-$ or $u-$channel. The correction therefore
increases in size beyond the $h-$pole, reaching a maximum close to the
relative size of the correction to the cross section for the $e^+e^-$ final
state at $M_1 \simeq 80$ GeV.

At larger values of $M_1$ additional channels begin to become accessible:
annihilation into $W^+W^-, \, ZZ, \, Zh$ and $hh$ final states becomes
possible at $M_1 \simeq 84$ GeV, $94$ GeV, $106$ GeV and $118$ GeV,
respectively. Since the gauge vertices are not corrected and in many diagrams
the Higgs vertex again only occurs linearly in the matrix element, the
corrections begin to decrease; however, the total annihilation cross section
is still dominated by $\ell^+ \ell^-$ final states.

At $M_1 \simeq 176$ GeV annihilation into $t \bar t$ opens up. Since this
cross section is not $P-$wave suppressed, it contributes significantly to the
total annihilation cross section, although the relevant matrix element is
suppressed either by the large stop masses (for the $t-$ and $u-$channel
diagrams) or the small $\lneut \lneut$ Higgs couplings (for the $s-$channel
Higgs exchange diagrams). The opening of this channel leads to a further
reduction of the relative size of the corrections.

For yet larger values of $M_1$, $s-$channel exchange of the heavy Higgs bosons
$H$ and $A$ begins to dominate the total cross section. Since the
corresponding matrix elements are again linear in the coupling we modify, the
relative size of the corrections continues to decrease until, near $M_1 = 300$
GeV, in rapid succession the $H^\pm W^\mp, \, Z A, \, ZH$ and finally $hA$ and
$hH$ final states become accessible. Many of these final states contribute
significantly to the total annihilation cross section; note that, unlike $hh$
and $\ell^+ \ell^-$ production, $hA$ production is possible from an $S-$wave
initial state, i.e. the cross section is not suppressed by a factor $v^2 =
0.09$ in our calculation. Moreover, final states with two Higgs bosons can be
produced via the exchange of higgsino--like states in the $t-$ or $u-$channel;
the corresponding diagrams are not suppressed by small couplings, and the
suppression ${\cal O}(M_1/\mu)$ by the heavy higgsino propagator is not very
severe for $M_1 \sim 300$ GeV. Since these diagrams involve two $\tilde \chi
\tilde \chi$ Higgs couplings, the relative size of the corrections begins to
increase again for $M_1 > 300$ GeV.

Finally, for $M_1 \geq 410$ GeV the wino component of $\lneut$ begins to grow
quickly. This leads to a rapid increase of the annihilation cross section into
$W^+W^-$ pairs, mostly via $t-$ and $u-$channel $\tilde \chi^+_1$ exchange
diagrams. Since we do not modify any of the relevant couplings, the
corrections quickly decrease, and remain very small once $|N_{12}| \simeq 1$.

The relative size of the correction to the predicted $\lneut$ relic density is
shown in Fig.~\ref{fig:m1}b. For $M_1 \leq 350$ GeV it is essentially the
mirror image of the correction to the total annihilation cross section, which
we just discussed, because to good approximation the relic density is
inversely proportional to the total annihilation cross section. There are some
minor differences, because the evaluation of the relic density involves
thermal averaging over a range of velocities, followed by integration over a
range of temperatures, whereas the results of the cross section had been
computed for a fixed relative $\lneut$ velocity, as mentioned above. As a
result, the relative corrections to the relic density are slightly smaller in
magnitude than those to the total annihilation cross section. This can be
understood from the usual (semi--)analytical solution \cite{ana_sol} of the
Boltzmann equation determining the relic density. For a $P-$wave initial
state, the relic density is inversely proportional to the product of the
square of the decoupling temperature and the annihilation cross section. Our
positive corrections increase the annihilation cross section, but also
slightly decrease the decoupling temperature; because the latter depends only
logarithmically on the annihilation cross section, the predicted relic density
is indeed reduced, but by a slightly smaller amount than if the decoupling
temperature was held fixed. The rapid change of the total annihilation cross
section, and resulting change of decoupling temperature, also leads to a small
kink at $M_1 = 256$ GeV in the relative corrections to the relic density,
right at the $H, A$ exchange poles.

For $M_1 \geq 350$ GeV, co--annihilation with the wino--like states $\tilde
\chi_2^0$ and $\tilde \chi_1^\pm$ begins to be important. For $M_1 > 410$ GeV,
$\lneut$ itself is wino--like. A small discontinuity occurs at the cross--over
value of $M_1$, since in our on--shell renormalization of the $\tilde \chi$
sector the mass of the most bino--like state is an input, i.e. is not
corrected, while the mass of the most wino--like neutralino does receive
(small) corrections. For $M_1 > 410$ GeV we therefore not only correct certain
$\lneut$ couplings, but also its mass.\footnote{The discontinuity in $\Delta
  N_{11}$ shown in Fig.~\ref{fig:m1}a has the same origin.} 

In this region, co--annihilation with the (now bino--like) $\tilde \chi_2^0$
no longer plays much of a role, but co--annihilation with $\tilde \chi_1^\pm$
remains very important. This can lead to $W^\pm Z$ final states, which are not
affected by our corrections, as well as $f \bar f'$ final states. The latter
can be produced both via $W^\pm$ exchange in the $s-$channel and by $\tilde f,
\tilde f'$ exchange in the $t-$ or $u-$channel. Squark exchange diagrams are
strongly suppressed by the large squark masses in our benchmark point. For
leptonic final states we find strong cancellation between $W$ and $\tilde
\ell$ exchange diagrams, which enhance the importance of the corrections to
the $\lneut \ell \tilde \ell$ as well as $\tilde \chi_1^+ \ell \tilde \ell'$
vertices. However, since most relevant (co--)annihilation cross sections are
basically not sensitive to our corrections, the total size of the corrections
to the $\lneut$ relic density remains below 1\% here. Note that the correction
to the relic density here has the same sign as the correction to the $\lneut$
annihilation cross section shown in Fig.~\ref{fig:m1}d, since co--annihilation
processes are more important than $\lneut \lneut$ annihilation reactions. Our
corrections lead to an even more perfect cancellation between $W^\pm$ and
$\tilde \ell$ exchange contributions to the production of leptonic final
states in $\lneut \tilde \chi_1^\pm$ co--annihilation, thereby slightly
increasing the predicted $\lneut$ relic density.

Recall that all quantities in Fig.~\ref{fig:m1} are plotted against
the $\overline{\rm DR}$ parameter $M_1$. The conversion of this, and
other input $\overline{\rm DR}$ parameters in the chargino and
neutralino mass matrices, into on--shell parameters made by
\texttt{SuSpect} is exact only in the limit of vanishing mixing
between the $\tilde \chi$ states. According to
ref.\cite{Pierce:1996zz} this reproduces the exact on--shell masses
quite accurately. This is consistent with our finding that one--loop
corrections to $\tilde \chi$ masses are small, at least in our version
of the on--shell scheme. However, in the region of strong bino--wino
mixing, where bino--wino co--annihilation is important, a small change
of parameters in the $\tilde \chi$ mass matrices can lead to
relatively large changes of mixing angles. In particular, using
running $\overline{\rm DR}$ values of $M_W, \, M_Z$ and $\theta_W$ in
the off--diagonal elements of the $\tilde \chi$ mass matrices, as in
the original \texttt{SuSpect}, leads to results for $N_{12}$ and
$N_{22}$ which differ by up to $5\%$ from those we obtain by using
on--shell values of $M_W, \, M_Z$ and $\theta_W$ everywhere, whereas
the values of the $\tilde \chi$ masses agree to within a fraction of a
percent between the two calculations.

It might seem peculiar that the original \texttt{SuSpect} uses
running, scale--dependent parameters in what are meant to be on--shell
mass matrices. However, while on--shell masses are well defined as the
poles of the (real parts of the) corresponding propagators, there is
no correspondingly simple definition of a (non--diagonal) ``on--shell
mass {\em matrix}''. Indeed, we could have performed an on--shell
renormalization of the $\tilde \chi$ sector where the off-diagonal
elements of the tree--level mass matrices were defined in terms of
running $M_W, \, M_Z$ and $\theta_W$; of course, (the finite parts of)
certain counterterms would then differ from their values in our
calculation. If one insists on using $\overline{\rm DR}$ input
parameters, as in the supersymmetric Les Houches accord \cite{SLHA},
the most consistent treatment is to use $\overline{\rm DR}$
renormalization throughout. On the other hand, if one insists on using
on--shell renormalization, as we (and most other calculations of
quantum corrections in supersymmetric theories) do, it seems more
consistent to use the masses of the three input states as free
parameters, rather than some $\overline{\rm DR}$ masses. If we compare
results of our modified \texttt{SuSpect} with the original one for
fixed values of the three input masses (e.g. of $\tilde \chi_1^\pm, \,
\tilde \chi_1^0$ and $\tilde \chi_3^0$ for $M_1 \leq 410$ GeV), rather
than for fixed $M_1, \, M_2$ and $\mu$, the two calculations also give
very similar $\tilde \chi$ gaugino components of the two lightest
neutralinos.\footnote{The higgsino components of these gaugino--like
  states, and the gaugino components of the higgsino--like states,
  still differ between these two tree--level calculations, but these
  have little effect on the relic density in this region of parameter
  space. Presumably these differences become very small if one
  compares one--loop calculations, since the two ``tree--level''
  neutralino mass matrices only differ by ${\cal O}(\alpha)$ terms.}
This does not solve the problem of fixing the definition of on--shell
mass matrices starting from $\overline{\rm DR}$ input parameters; this
obverservation is nevertheless reassuring, since it indicates that the
precise definition of the on--shell mass matrix may not matter very
much once actual measurements are used to fix the parameters of the
theory.

Finally, we mention that the $\lneut$ relic density attains the desired value
$\Omega_{\lneut} h^2 \simeq 0.11$ for $M_1 \simeq 230$ GeV, $\simeq 265$ GeV
(to either sides of the $H,A$ poles) and $\simeq 375$ GeV (where the wino
component of $\lneut$ as well as co--annihilation with the wino--like states
$\tilde \chi_2^0$ and $\tilde \chi_1^\pm$ become significant). For
$230 \ {\rm GeV} < M_1 < 265 \ {\rm GeV}$ as well as for $M_1 > 380$
GeV the LSP could contribute a sub--dominant component of the overall
cosmological Dark Matter.

\subsection{$M_2$ dependence}

\begin{figure}[!ht]
\epsfig{file=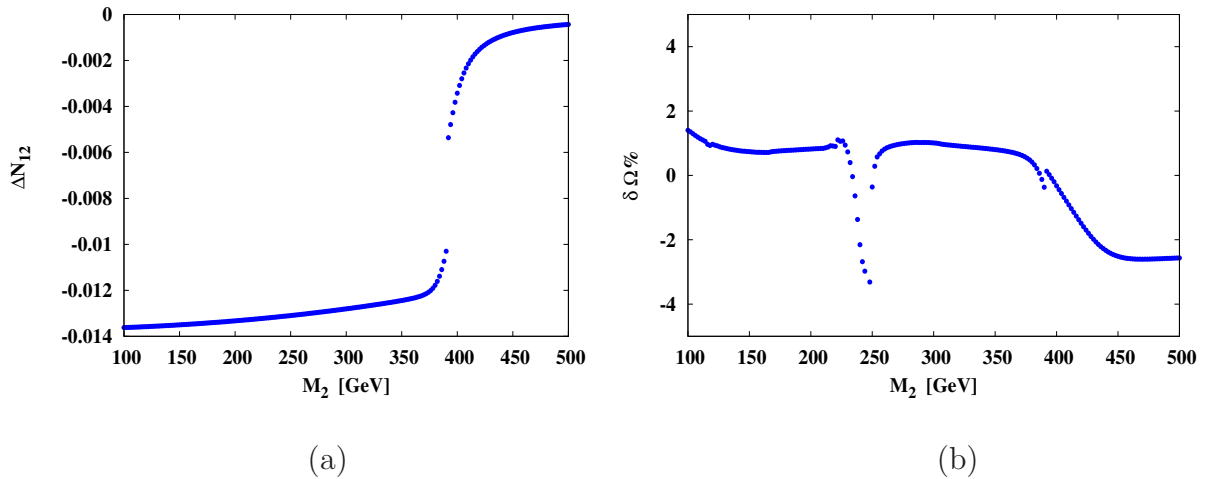, width=\linewidth, height=6 cm} 
$~~~~~~~~~~~~~~~~~~~~~~~~~~~~~~ \text{(a)}  \hspace{7.75cm} \text{(b)}$
\caption{(a) $\Delta N_{12}$, as given by eq.(\ref{eq:dUVN2}) and (b)
  the relative size of radiative corrections to the relic density of
  $\tilde \chi^{0}_{1}$, as defined in eq.(\ref{delsig}), against
  $M_2$, for our benchmark scenario (\ref{bench}) with $M_1 = 400$
  GeV.}
\label{fig:m2} 
\end{figure}  

In Fig.~\ref{fig:m2}a, b we plot the dependence of the absolute
correction to the wino coupling of $\lneut$ and the relative
correction to the $\lneut$ relic density, respectively, on the wino
mass parameter $M_2$ for bino mass parameter $M_1 = 400 $ GeV. Note
that $N_{12} < 0$ here, i.e. the corrections increase the wino
coupling $|N_{12}|$. The bulk of these corrections is RG--like,
i.e. can be attributed to the running of the $SU(2)$ component of the
$\tilde \chi f \tilde f$ ``gauge'' coupling. At first glance it might
be surprising that the size of these ${\cal O}(g^2)$ corrections is
somewhat smaller than the ${\cal O}(g_Y^2)$ correction shown in
Fig.~\ref{fig:m1}a, even though $g^2 \simeq 3.3 g_Y^2$. The reason for
this is that in the on--shell scheme the wino coupling receives a
sizable negative ``constant'' correction (which does not grow
logarithmically with the sfermion masses): the rather small value of
$\sin^2 \theta_W \simeq 0.222$ leads to a tree--level $SU(2)$ gauge
coupling $g = e/\sin \theta_W$ that is somewhat larger than the
coupling used in standard \texttt{micrOMEGAs}, where $\sin^2\theta_W =
0.231$. The total wino coupling $g (N_{12} + \Delta N_{12})$ is
independent of the choice of renormalization scheme.

Results for $\Delta U_{11}$ and $\Delta V_{11}$, which describe the
wino couplings of the lighter chargino, are very similar in magnitude
to $\Delta N_{12}$; these chargino couplings contribute to certain
co--annihilation reactions.

At $M_2 \sim 390$ GeV there is a discontinuity in $\Delta N_{12}$. This is
because the dominantly wino--like neutralino no longer remains the lightest
one, the bino--like one becomes the lightest state; we had found an analogous
discontinuity in Fig.~\ref{fig:m1}. As $M_2$ is increased beyond this value,
$|\Delta N_{12}|$ quickly decreases, simply because $|N_{12}|$ decreases as
$\lneut$ becomes increasingly bino--like.

In most of the region with wino--like $\lneut$ that satisfies the LEP
constraint $M_2 \geq 110$ GeV \cite{PDG} the corrections to the $\lneut$ relic
density shown in Fig.~\ref{fig:m2}b are quite small and positive. Here both
$\lneut$ annihilation into $W^+W^-$ pairs, and various $\lneut \tilde
\chi_1^\pm$ co--annihilation reactions are important. As in Fig.~\ref{fig:m1}b
for $M_1 > 400$ GeV, co--annihilation into leptonic final states is strongly
suppressed by destructive interference between $W^\pm$ and slepton exchange
diagrams. The increase of the magnitude $\lneut$ and $\tilde \chi_1^\pm$
couplings to $SU(2)$ doublet (s)leptons further strengthens this cancellation;
since this final state is in any case subdominant, the corrections increase
the predicted relic density only slightly.

Within the region where $\lneut$ is wino--like, the corrections to the
relic density become significant only in the vicinity of the
$s-$channel $A, \, H$ poles at $M_2 \simeq 250$ GeV. The couplings of
these Higgs bosons are essentially proportional to $N_{12}$ here,
$N_{11}$ being very small. Since our corrections increase $|N_{12}|$,
as noted above, they increase the Higgs exchange contributions to the
annihilation cross section, and thereby reduce the relic
density. Recall also that our tree--level $SU(2)$ gauge coupling is
about $1.5\%$ larger than that used by \texttt{micrOMEGAs}; this also
contributes to the corrections as defined in eq.(\ref{delsig}). Note
that there is also an $H^\pm$ exchange pole in $\lneut \tilde
\chi_1^\pm$ co--annihilation, which also contributes in this range of
$M_2$ since the masses of $A, \, H$ and $H^\pm$ are within a few
hundred MeV of each other. Since, as noted above, the wino coupling of
$\tilde \chi_1^\pm$ is enhanced similarly as that of $\lneut$, our
corrections also enhance these contributions to the co--annihilation
cross section. However, in both cases the Higgs couplings to $\tilde
\chi$ currents are suppressed by the small higgsino components of
$\lneut$ and $\tilde \chi_1^\pm$; hence Higgs exchange in the
$s-$channel is important only for a narrow range of
$M_2$.\footnote{Since a wino--like $\lneut$ isn't an input state in
  our on--shell renormalization of the $\tilde \chi$ sector, its mass
  receives non--vanishing one--loop corrections in our
  calculation. However, they amount to less than 200 MeV in magnitude,
  significantly smaller than the total decay width of the heavy Higgs
  bosons, which amounts to $\sim 1$ GeV in our benchmark
  scenario. These corrections to the $\lneut$ mass are therefore
  essentially irrelevant near the heavy Higgs poles, but lead to
  sizable changes of the relic density near the $h$ pole; however, the
  lower bound on the $\tilde \chi_1^\pm$ mass from LEP experiments
  \cite{PDG} excludes the region of parameter space where a wino--like
  $\lneut$ has mass near $m_h/2$.}

Finally, for $M_2 > 400$ GeV $\lneut$ becomes increasingly bino--like, as
mentioned above. Here the corrections to the bino coupling, $\Delta N_{11}$,
are of similar size as those shown in Fig.~\ref{fig:m1}a for $M_1 < 400$
GeV. In the transition region, co--annihilation with $\tilde \chi_2^0$ and
$\tilde \chi_1^\pm$ is important. For yet larger values of $M_2$, near the end
of the range shown in Fig.~\ref{fig:m2}, co--annihilation with these
wino--like states is suppressed by the large bino--wino mass splitting, and
annihilation into $hA, ZH$ and leptonic final states dominates. Since our
corrections increase the bino coupling of $\lneut$ by almost $2\%$, they
increase many of the relevant cross sections, thereby reducing the predicted
relic density by several percent.

In standard cosmology, the $\lneut$ relic density reaches its desired
value for $M_2 \simeq 425$ GeV, where $\lneut$ still has significant
wino--component\footnote{Of approximately equal size, but opposite
  sign, of a photino state.}, and co--annihilation with $\tilde
\chi_2^0$ and $\tilde \chi_1^\pm$ is still significant. Our
corrections reduce the relic density by about $2\%$ at this point. For
$M_2 < 400$ GeV, $\lneut$ could only contribute less than $10\%$ of
the desired Dark Matter density.

\subsection{$\mu$ dependence}
\setcounter{footnote}{0}

Having discussed our corrections for bino-- and wino--like $\lneut$, we now
consider a region of parameter space where $\lneut$ is higgsino--like, by
fixing $M_1 = 400$ GeV, $M_2 = 600$ GeV and varying $\mu$, chosing $\mu >
0$. Note that there are two higgsino--like neutralinos. One of these states
shows significantly stronger higgsino--gaugino mixing, which simultaneously
reduces its higgsino content and its mass. Since the other, heavier
higgsino--like state also has (even) larger higgsino content, its mass is one
of the input masses in our version of the on--shell renormalization of the
$\tilde \chi$ sector. This means that the $\lneut$ mass does receive radiative
corrections in our scheme, which can amount to several GeV. At the same time,
the gaugino couplings of $\lneut$ receive corrections whose {\em relative}
size is similar to those shown in figs.~\ref{fig:m1}a and \ref{fig:m2}a for
the bino-- and wino--coupling, respectively; recall, however, that the
bino-- and wino--couplings of higgsino--like states also receive
corrections that can be interpreted as corrections to the
off--diagonal entries of the $\tilde \chi$ mass matrices, rather than
corrections that come from the running of the ``gauge'' couplings of
the $\tilde \chi$ states.

\begin{figure}[!ht]
\epsfig{file=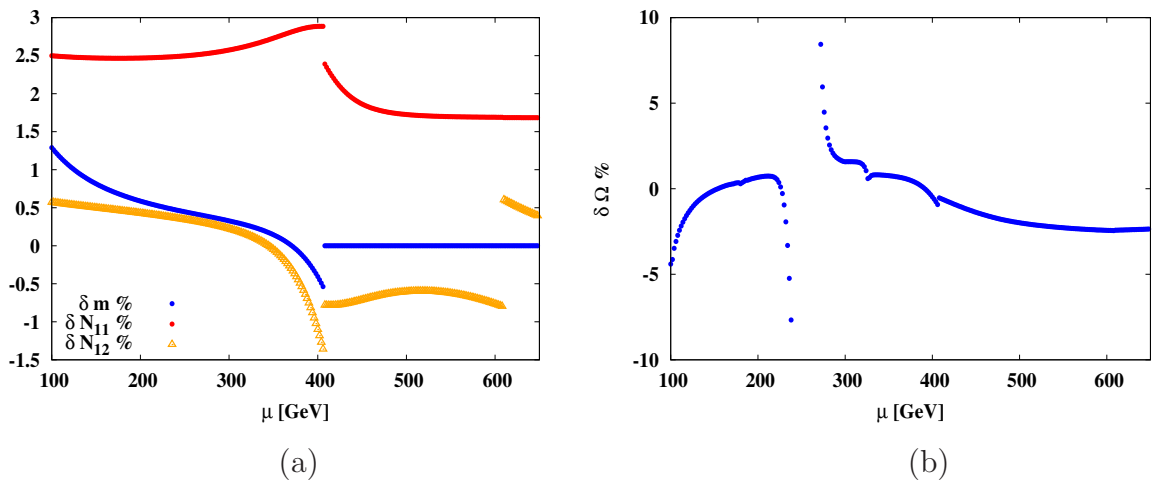, width=\linewidth, height=6 cm} 
$~~~~~~~~~~~~~~~~~~~~~~~~~~~~~~ \text{(a)}  \hspace{7.75cm} \text{(b)}$
\caption{Relative size of the corrections to (a) the mass and gaugino
  couplings of $\lneut$, and (b) to the predicted $\lneut$ relic
  density against $\mu$, for our benchmark point (\ref{bench}) with
  $M_1 = 400$ GeV and $M_2 = 600$ GeV.}
\label{fig:mu} 
\end{figure}

In Fig.~\ref{fig:mu}a we plot the relative size of the corrections to
the mass and to the gaugino entries $N_{11}$ and $N_{12}$. For $\mu
\leq 300$ GeV the absolute correction to the $\lneut$ mass is
approximately $1.3$ GeV, nearly independent of $\mu$, leading to a
falling relative size of the correction with increasing $\mu$. For
even larger values of $\mu$ the bino--component of $\lneut$ begins to
become sizable; this leads to a rapid decrease of the correction to
the mass, which turns negative at $\mu \simeq 370$ GeV. Note, however,
that the correction remains small in magnitude even at $\mu = M_1 =
400$ GeV, where other versions of the on--shell scheme are badly
behaved \cite{CDKX}. For yet larger values of $\mu$, $\lneut$ is
bino--like; its mass is then an input, and does not receive any
corrections.

Much of the correction to the relic density shown in
Fig.~\ref{fig:mu}b can be explained through this correction to the
$\lneut$ mass. Note that higgsino--like neutralinos strongly
(co--)annihilate into final states containing two gauge
bosons;\footnote{In principle $W^+W^-$ and $ZZ$ final states can be
  produced via the $s-$channel exchange of the CP--even Higgs bosons,
  whose couplings to two neutralinos or two charginos do receive some
  corrections. However, these contributions are suppressed by the
  small size of the gaugino components of the (co--)annihilating
  states; moreover, since $m_A^2 \gg M_Z^2$, the couplings of the
  heavy CP--even Higgs boson $H$ to two gauge bosons is very small. In
  practice our corrections to $\lneut$ couplings therefore do not
  significantly modify the (co--)annihilation cross sections into two
  gauge bosons.} co--annihilation into $f \bar f$ final states via
$s-$channel exchange of a gauge boson is also important, while $t-$
and $u-$channel sfermion exchange diagrams are suppressed by large
sfermion masses (in case of squarks) and/or small Yukawa couplings of
the corresponding fermions (in case of first and second generation
sfermions). As a result, for most of parameter space none of these
cross sections is affected significantly by our corrections to
$\lneut$ couplings.\footnote{The corrections to the ``top Yukawa''
  coupling of $\lneut$, which is $\propto \Delta N_{14}$, does play a
  role in the tiny kink visible at the $t \bar t$ threshold, $\mu
  \simeq 180$ GeV.}

The total correction to the relic density can nevertheless be
sizable. This is most obvious in the vicinity of the heavy Higgs
poles, where the positive shift of the $\lneut$ mass is important,
since it slightly exceeds the width of the heavy Higgs bosons. It
therefore significantly increases the cross section below the
resonance, where the correction reduces the mass gap to the resonance,
but reduces it above the resonance, where the corrections move
$\lneut$ further away from the resonance. The positive correction to
the $\lneut$ mass also increases the annihilation cross section for
rather light $\lneut$, close to the $W^+W^-$ and $ZZ$ thresholds where
the corresponding cross sections are still significantly phase space
suppressed. On the other hand, away from poles and thresholds
increasing the $\lneut$ mass slightly decreases the annihilation cross
section, which basically scales like $1/m^2_{\lneut}$ in this region.

In principle one can chose a renormalization scheme where the LSP mass
is used as input in the on--shell renormalization of the $\tilde \chi$
sector. However, such a scheme would lead to quite large corrections
in some regions of parameter space \cite{CDKX}. Moreover,
co--annihilation reactions play a significant role in determining the
relic density of a higgsino--like LSP. If the LSP mass is kept fixed,
there will be significant loop corrections to the masses of the other
higgsino--like states, $m_{\tilde \chi_2^0}$ and $m_{\tilde
  \chi_1^\pm}$;\footnote{A renormalization scheme were the masses of
  all three higgsino--like states were used as inputs would be very
  badly behaved indeed, since then no input mass would have
  significant dependence on $M_1$ or $M_2$, leading to very large
  counterterms for these parameters \cite{CDKX}.} as a result, there
would be large corrections to co--annihilation cross sections
proceeding through exchange of a heavy (neutral or charged) Higgs
boson. The total size of the correction to the relic density near the
heavy Higgs poles would then still be of similar size as in our scheme.

In the vicinity of the Higgs poles, our corrections to the wino-- and
bino--couplings of $\lneut$ (and, in co--annihilation reactions, of
$\tilde \chi_2^0$, as well as the wino couplings of $\tilde
\chi_1^\pm$) do play a role. Note that the wino-- and bino--components
of $\lneut$ have opposite sign in this region of parameter space,
which means that they add constructively to the $\lneut \lneut (H,A)$
couplings. We see in Fig.~\ref{fig:mu}a that in the region of the
heavy Higgs poles, the relative corrections to $N_{11}$ and $N_{12}$
are both positive, leading to an increase of the magnitudes of the
$\lneut \lneut (H,A)$ couplings. For larger $\mu$ the correction to
$N_{12}$ turns negative, helping to reduce the overall size of the
corrections to the relic density in that region of parameter space.

The corrections to (the gaugino part of) Higgs couplings to $\tilde
\chi$ currents also play a role in the (co-)annihilation cross
sections into one massive gauge boson and one heavy Higgs boson, which
open around $\mu = 310$ GeV, as well into the light Higgs boson $h$
and one of the heavy Higgs bosons $A, H, H^\pm$, which become
accessible for $\mu > 322$ GeV; the former mostly proceed via the
exchange of a heavy Higgs boson in the $s-$channel, while the latter
also receive significant contributions from the exchange a
gaugino--like $\tilde\chi$ states in the $t-$ or $u-$channel. With the
exception of the region with strongest bino--higgsino mixing, $\mu
\simeq 400$ GeV, where the corrections to $N_{12}$ are relatively
large and negative, all relevant Higgs couplings are enhanced by our
correction, thereby increasing the corresponding cross sections,
whereas the increase of the mass still tends to reduce the cross
section due to the proximity of the $H,A$ poles. These two effects
tend to cancel, reducing the overall size of the correction to the
relic density to $\leq 1.5\%$ for $300 \ {\rm GeV} \leq \mu \leq 400$
GeV.

For yet larger $\mu$, $\lneut$ becomes bino--like, as noted above. The
absolute size of the correction to $\Omega_{\lneut}$ then increases
with increasing $\mu$, as co--annihilation reactions, which are little
affected by our corrections, become less important. Since the
corrections increase the bino coupling of $\lneut$, they reduce the
relic density. The correction is somewhat smaller than in most of
Fig.~\ref{fig:m1}d since even at $\mu = 600$ GeV final states
containing one massive gauge boson and one Higgs boson, as well as
diagrams with Higgs exchange in the $s-$channel, play significant
roles; in both cases only one vertex gets corrected. Note finally that
the correction is perfectly well--behaved at $\mu = M_2 = 600$ GeV, as
advertised (and in contrast to the scheme used in
\cite{Guasch:2002ez}).

Within standard cosmology, the $\lneut$ relic density reaches its
desired value at $\mu \simeq 460$ GeV, in the region of strong
higgsino--bino mixing, where corrections to $\Omega_{\lneut}$ are
negative. For smaller values of $\mu$, $\lneut$ would only contribute
a fraction of the total Dark Matter density. 

\subsection{Co-annihilation with $\tilde{\tau_1}$}
\setcounter{footnote}{0}

If the mass difference between the LSP and the next--lightest
superparticle (NLSP) is sufficiently small, co--annihilation processes
play a role in the determination of the LSP relic density
\cite{Griest:1990kh}. This is generic for wino-- as well as
higgsino--like LSP, where $\tilde \chi_1^\pm$ is automatically close
in mass to $\lneut$; in case of higgsino--like LSP, $\tilde \chi_2^0$
is also nearby. In this subsection we study co--annihilation of a
bino--like LSP with the lightest stau state $\tilde \tau_1$. This is
well motivated even within the framework of the minimal supergravity
(mSUGRA) model \cite{Ellis:1998kh}. The effect of co--annihilation can
be described quite well \cite{Griest:1990kh} by introducing an
effective annihilation cross section, which is essentially the sum of
the usual LSP (self--)annihilation cross section and the $\lneut
\tilde \tau_1$ and $\tilde \tau_1 \tilde \tau_1$ annihilation cross
sections, where the latter are suppressed by one and two powers,
respectively, of the Boltzmann factor $\exp[-T_F/(m_{\tilde \tau_1} -
m_{\lneut})]$. Since the freeze--out temperature $T_F$ is basically
proportional to $m_{\lneut}$, the Boltzmann suppression factor depends
on the {\em relative} mass splitting between the LSP and the lighter
stau.

\begin{figure}[!ht]
\epsfig{file=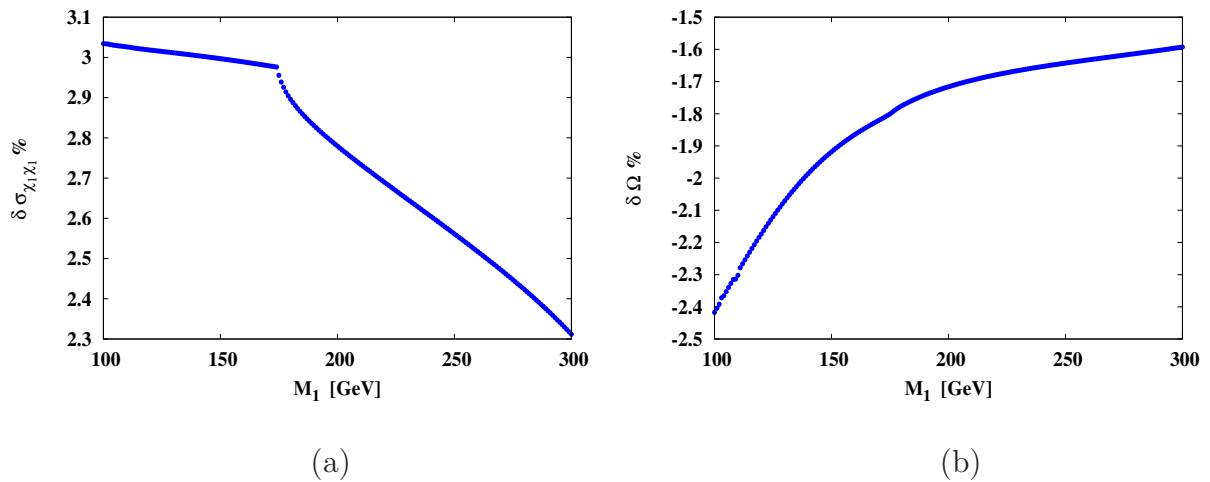, width=\linewidth, height=6 cm} 
$~~~~~~~~~~~~~~~~~~~~~~~~~~~~~~ \text{(a)}  \hspace{7.75cm} \text{(b)}$
\caption{(a) Correction to the $\lneut$ annihilation cross section and (b)
  radiative correction to the relic density of $\lneut$ (in \%)
  against $M_1$. Results are for $M_2 = 0.4$ TeV, $\mu = 0.6$ TeV,
  $m_A = 1.0$ TeV, $\tan\beta = 5$ and $m_{\tilde \tau_R} = M_1 + 5$
  GeV; the remaining parameters are as in our benchmark scenario
  (\ref{bench}).} 
\label{fig:coann} 
\end{figure}

We have assumed $\mu$ = 600 GeV and $M_2$ = 400 GeV. All other input
parameters are as described in (\ref{bench}), except that we have used
$m_A = 1.0$ TeV\footnote{We have increased $m_A$ in order to avoid the
  heavy Higgs poles, where \texttt{micrOMEGAs} (version 2.2.CPC.i)
  produces a spurious divergence in the co--annihilation cross
  section. This occurs because $\tilde \tau_1 \rightarrow \tau +
  \lneut$ can occur on--shell. As a result, $\tilde \tau_1 + \lneut
  \rightarrow \tau + (A,H)$ can proceed through ``exchange'' of an
  on--shell $\lneut$, giving an infinity in the (differential) cross
  section; note that $\lneut$ is stable, and hence has vanishing
  width. This divergence is not physical. In the present context, the
  very fast reactions $\tilde \tau_1 \leftrightarrow \lneut + \tau$
  help to maintain the relative equilibrium between the $\lneut$ and
  $\tilde \tau_1$ densities at temperatures well below $T_F$. In the
  formalism of ref.\cite{Griest:1990kh} these fast reactions are
  integrated out to arrive at the effective Boltzmann equation for the
  $\lneut$ density. The removal of this unphysical divergence is a
  separate issue not related to the radiative corrections we are
  interested in, so we simply avoid it by increasing the masses of the
  heavy Higgs bosons.}, $m_{\tilde \tau_R} = M_1 + 5$ GeV and $\tan
\beta = 5$. Note that the physical $\tilde \tau_1$ mass is increased
by the ``$D-$term'' contribution $-\sin^2 \theta_W \cos(2\beta) M_Z^2
> 0$, and reduced by $\tilde \tau_L - \tilde \tau_R$ mixing. For the
given choice of parameters, we find a $\tilde \tau_1 - \lneut$ mass
splitting of about 14 GeV, almost independent of $M_1$ in the range
shown in Fig.~\ref{fig:coann}.

In Fig.~\ref{fig:coann} (a), the correction to the $\lneut$
annihilation cross section is shown. Since $\tilde \tau_1$ is much
lighter than all other sfermions in the scenario considered here,
$\lneut$ pairs dominantly annihilate into $\tau^+\tau^-$ pairs via
$\tilde \tau_1$ exchange in the $t-$ or $u-$channel. Both vertices in
these diagrams receive significant positive corrections, leading to an
enhanced $\lneut$ annihilation cross section.

At $M_1 \sim 175$ GeV the relative size of the correction begins to
diminish somewhat. At this point annihilation to a $t \bar t$ pair
opens. Increasing $M_1$ along with $m_{\tilde \tau_R}$ reduces the
annihilation cross section into $\tau^+\tau^-$ pairs roughly like the
inverse square of the $\lneut$ mass, but increases several other cross
sections. In particular, since $\mu$ is kept fixed, increasing $M_1$
increases higgsino--gaugino mixing, which leads to enhanced cross
sections for $W^+W^-, \, ZZ$ as well as $Z h$ final states. In some of
these channels, as well as for the $t \bar t$ channel, there is
destructive interference between $s-$channel Higgs exchange and $t-$
and $u-$channel diagrams. The corrections to the annihilation cross
sections into $W^+W^-$ and $t \bar t$ final states are therefore
negative (about $-0.5\%$ and $-3\%$, respectively).  This leads to a
reduction of the relative size of the correction to the total $\lneut$
annihilation cross section.

The right frame of Fig.~\ref{fig:coann} shows that the correction to
the $\lneut$ relic density is significantly smaller in magnitude than
that to the $\lneut$ annihilation cross section. This is mostly due to
co--annihilation contributions. Recall that the importance of these
contributions depends on the {\em relative} $\tilde \tau_1 - \lneut$
mass splitting; since the absolute mass splitting remains
approximately constant over the entire range of $M_1$ shown, the
relative mass splitting decreases approximately $\propto 1/M_1$. As a
result, co--annihilation becomes more important at larger $M_1$.

The most important co--annihilation reactions are $\lneut \tilde
\tau_1^\pm \rightarrow \tau^\pm + (\gamma, \, Z)$, where only one
vertex is corrected, as well as $\tilde \tau_1^+ \tilde \tau_1^-
\rightarrow \gamma \gamma, \ \gamma Z$, which proceed solely through
(true) gauge interactions, which we do not correct. However, $\tilde
\tau_1^\pm \tilde \tau_1^\pm \rightarrow \tau^\pm \tau^\pm$ also
contributes significantly, where again both vertices of the dominant
$\tilde \chi^0$ exchange diagrams get corrected. Nevertheless, our
corrections affect the co--annihilation cross sections much less than
the $\lneut$ annihilation cross section. Together with the increased
importance of these co--annihilation contributions this explains the
decline of the relative size of the corrections with increasing
$M_1$. In the case at hand, co--annihilation is dominant for $M_1 \geq
150$ GeV. For much larger values of $M_1$, the Boltzmann suppression
factors are close to unity, and the relative size of the corrections
stays approximately constant until one approaches the heavy Higgs
poles (not shown).

Within standard cosmology, in the parameter space shown in
Fig.~\ref{fig:coann}, the $\lneut$ relic density is three to four times
higher than the desired value. This could be corrected by also
reducing the masses of some other sleptons, and/or further reducing
$m_{\tilde \tau_1}$. In both cases the importance of co--annihilation
contributions would increase, so that the relative corrections to the
relic density would be similar to that in the high$-M_1$ region of
Fig.~\ref{fig:coann}.

\section{Conclusions}
\label{sec:Conc}
\setcounter{footnote}{0}

In this paper we have studied the impact of the effective couplings,
as introduced in \cite{Guasch:2002ez}, and of corrections to the
neutralino and chargino masses, in the scheme introduced in
ref.\cite{CDKX}, on the calculation of the cosmological relic density
of the lightest neutralino in the MSSM. 

In Sec.~2 we analyzed the nature of the corrections captured by these
effective couplings. We pointed out that they not only describe the
difference in the running of true gauge couplings and gaugino
couplings, but also corrections to the entries of the chargino and
neutralino mass matrices that mix higgsino and gaugino
states. Technically, these effective couplings include $\tilde \chi$
two--point function corrections plus associated counterterms. These
corrections are thus enhanced by multiplicity factors, since all
(s)fermions contribute similarly to the gaugino two--point
functions. It has also been known for some time that the magnitude of
these corrections increases logarithmically with the mass of heavy
sfermions \cite{Hikasa:1995bw}. There is thus some reason to believe
that they capture a large fraction of all electroweak radiative
corrections; in case of $\tilde\chi_2^0 \rightarrow f \bar f \lneut$
decays this has been shown to be the case \cite{dhx}.

In Sec.~3 we briefly described how we incorporated these corrections
into the \texttt{micrOMEGAs} program package \cite{Belanger:2008sj}.
The corrections apply to both the ``gauge'' and the ``Yukawa''
components of $\tilde \chi \tilde f f$ couplings, as well as to the
``gaugino'' components of Higgs couplings to $\tilde \chi$
currents. It should be noted that we use on--shell renormalization in
the electroweak sector, whereas \texttt{micrOMEGAs} uses
$\overline{\rm MS}$ inputs for the QED coupling and the weak mixing
angle. We nevertheless compare with \texttt{micrOMEGAs} tree--level
predictions when illustrating the effect of the corrections, see
eq.(\ref{delsig}).\footnote{The corrected couplings $g(N+ \Delta N),
  \, g(U+ \delta U), \, g(V + \Delta V)$ are scheme--independent up to
  two--loop terms; however, the tree--level couplings $gN$ etc. are
  scheme--dependent already at one--loop order. We use
  \texttt{micrOMEGAs} choices here since this is currently the
  industry standard for LSP relic density calculations.}

Numerical results are shown in Sec.~4. We found that the corrections
to $\sigma(\lneut \lneut \rightarrow e^+e^-)$ can reach $+15\%$ for
wino--like $\lneut$. However, in this case $\lneut$ dominantly
annihilates into $W^+W^-$ final states, which are relatively little
affected by our corrections (only via $s-$channel Higgs exchange
diagrams). The corrections to the {\em total} $\lneut$ annihilation
cross section due to the effective couplings therefore never exceed
$4\%$ for our choice of squark masses near $1.5$ TeV, i.e. not far
beyond the current LHC lower bound. The correction to the $\lneut$
mass, which never exceed a couple of percent in our scheme, can lead
to significantly larger corrections to the total annihilation cross
section in the vicinity of Higgs poles. These mass corrections were
shown to be especially important for higgsino--like LSP. In all cases
the relative corrections to the relic density are slightly smaller
than those to the (effective) annihilation cross section, since an
increase of the latter also reduces the temperature at which $\lneut$
decouples from the thermal plasma. 

It should be noted that the corrections to the relic density are
frequently comparable to, or larger than, the present observational
uncertainty on the total Dark Matter density (in the framework of
standard cosmology); this uncertainty is expected to be reduced
significantly as soon as the PLANCK collaboration completes the
analysis of their data on the cosmic microwave anisotropies. 

Note also that the effective couplings as defined in
\cite{Guasch:2002ez} include certain diagrams contributing to a
complete one--loop calculation of $\sigma(\lneut \lneut \rightarrow f
\bar f)$ {\em exactly}, namely the two--point function corrections on
the external $\lneut$ lines with fermion--sfermion loops. Their
inclusion into an ``effective tree--level'' calculation, as done here,
can therefore be understood as a first step towards a full electroweak
one--loop calculation \cite{Baro:2007em}, which is however
computationally far more complex. We use the same effective couplings
also when a $\tilde \chi$ is exchanged in the $t-$ or $u-$channel,
which contributes to sfermion co--annihilation; since here $\tilde
\chi$ is off--shell, the effective couplings include two--point
function corrections to its propagator only approximately, but the
relation with the exact calculation is still fairly straightforward.

The situation is a bit more complicated for the Higgs couplings to
$\tilde \chi$ currents. Here we correct only the gaugino part of the
coupling. In the limit of vanishing gaugino--higgsino mixing, this
corresponds to including only the gaugino to gaugino and higgsino to
gaugino two--point functions, but no higgsino to higgsino or gaugino
to higgsino diagrams. However, in realistic situations with
non--vanishing $\tilde \chi$ mixing this no longer corresponds to a
complete subclass of diagrams. Moreover, it has been shown some time
ago \cite{Drees:1996pk} that at least some corrections to $Z \lneut
\lneut$ couplings can be described in terms of neutralino two--point
function corrections. There might thus be further scope for extension
of the effective coupling scheme to Higgs couplings, and/or to
couplings to $W$ and $Z$ bosons to $\tilde \chi$ states. We plan to
investigate these issues in future.

\section*{Acknowledgment}
We thank K. Williams and D. Lopez--Val for numerous discussions.  We
also thank F. Boudjema for discussions on \texttt{micrOMEGAs} and
issues of gauge invariance. AC and SK acknowledge support from the
Bonn Cologne Graduate School of Physics and Astronomy. This work was
partially supported by the DFG Transregio TR33 ``The Dark Side of the
Universe'', and partly by the German Bundesministerium f\"ur Bildung
und Forschung (BMBF) under Contract No. 05HT6PDA. S.K. acknowledges
partial support from the European Union FP7 ITN INVISIBLES (Marie
Curie Actions, PITN- GA-2011- 289442).

\appendix

\section{Weak Sector} 
\label{Appendix:weak}

We apply on--shell renormalization of the electroweak sector, where $M_W$
and $M_Z$ are physical (pole) masses, and $\cos\theta_W = M_W/M_Z$. This gives
\cite{Marciano:1980pb}:
\bea \label{delta_SM}
\delta M_{W}^{2} & = & \Re\Sigma_{WW}(M_{W}^{2})\,; \nonumber \\
\delta M_{Z}^{2} & = & \Re\Sigma_{ZZ}(M_{Z}^{2})\,; \nonumber \\
\delta \cos \theta_W & = & \frac{M_W}{M_Z} \left(\frac{\delta M_W}{M_W}-
  \frac{\delta M_Z}{M_Z} \right)\,.
\eea
Here $\Sigma_{WW}$ and $\Sigma_{ZZ}$ are the transverse components of
the diagonal $W$ and $Z$ two--point functions in momentum space,
respectively; only the real parts of the loop functions should
be included as indicated by the $\Re$ symbols. Note that, while the
definitions of these three counterterms are formally as in the SM, in
the MSSM there are many new contributions to $\Sigma_{WW}$ and
$\Sigma_{ZZ}$ involving loops of superparticles and additional Higgs
bosons.

In our treatment we have to calculate the counterterms for $M_W, \, M_Z$ and
$\cos\theta_W$ twice: for the purpose of computing effective couplings, we
only include loops of matter fermions and sfermions, but for the correction to
neutralino and chargino masses \cite{CDKX} we include the full MSSM
contributions. 

The $SU(2)$ gauge coupling is given by $g = e/\sin\theta_W$, where the
$e$ is the QED couplings constants, the renormalization of which is
discussed in the following Appendix. The counterterm $\delta g$ is
then given by
\beq \label{delta_g}
\delta g = \frac{ \delta e} {\sin \theta_W} - g \frac {\delta \sin \theta_W}
       {\sin\theta_W} = \frac {\delta e} {\sin \theta_W} + g \frac {\cos
         \theta_W} {\sin^2 \theta_W} \delta \cos \theta_W\,,
\eeq
where $\delta \cos\theta_W$ has been given in eq.(\ref{delta_SM}), and $\delta
e = e(M_Z) \delta Z_e^{e(M_Z^2)}$ with $\delta Z_e^{e(M_Z^2)}$ given in
eq.(\ref{delta_Ze}) below. The counterterm $\delta g$ is needed only in the
calculation of effective couplings, so only matter (s)fermion loops are
included. 

Similarly, the counterterm $\delta \tan\theta_W$ needed in
eqs.(\ref{eq:dUVN2}) is to one--loop order given by
\beq \label{dtantw}
\delta \tan \theta_W = - \frac {\delta \cos \theta_W} {\sin \theta_W \cos^2
  \theta_W} \,.
\eeq

\section{Charge Renormalization}
\label{Appendix:charge}

The on--shell renormalization condition for the electric charge requires the
electric charge to be equal to the full $ee\gamma$ coupling for on--shell
external particles in the Thomson limit. This leads to
\beq
  e^{\rm bare}\rightarrow e(0) \left(1+\delta Z_e^{(0)}\right).
\eeq
The renormalization constant is adjusted to cancel the loop corrections to the
$ee\gamma$ vertex in this limit. This gives \cite{Denner:1991kt}:
\beq
\delta Z_e^{(0)} = \frac{1}{2}\frac{\partial}{\partial q^2} \Sigma_{\gamma
  \gamma} (q^2)\displaystyle |_{q^2=0} + \tan\theta_W\frac{\Sigma_{\gamma 
Z}(0)}{M^2_{Z}},
\eeq
where the $\Sigma$ again refer to the transverse part of the corresponding
two--point functions. We can thus identify the renormalized charge $e(0) =
\sqrt{4\pi\alpha(0)}$ where $\alpha(0) = 1/137.036\dots$ is the fine structure
constant, as defined in the Thomson limit. 

While pure on--shell renormalization of $e$ is perfectly fine in a theory of
photons and electrons, a problem arises in the presence of light quarks. The
counterterm $\delta Z_e^{(0)}$ receives contributions from loops involving a
light fermion $f$ that scale $\propto \log(m_f^2)$, leading to terms
proportional to $\alpha \log\left(m^2_f/s\right)$ in corrections to physical
processes at some energy scale $s$; these can be understood as describing the
running of the QED coupling. The problem is that the masses of the light
$u,d,s$ (and, to some extent, $c$) quarks aren't well known, and in fact are
difficult to define in the Thomson limit.

Fortunately this dependence on the light quark masses can be traded for a
dependence on experimentally measured cross sections for $e^+e^-$ annihilation
into hadronic final states. To that end, one writes
\begin{equation} \label{split}
\frac{\partial}{\partial q^2}\Sigma_{\gamma\gamma}^{\textrm{light~}f\textrm{~in
    loops}}(q^2)\displaystyle |_{q^2=0} = \Delta \alpha +
\frac{1}{M^2_{Z}}\Re\Sigma_{\gamma\gamma}^{\textrm{light }f\textrm{ in
    loops}}(M^2_{Z}) \, ,
\end{equation}
where $\Delta \alpha $ is a finite quantity; moreover, the second term in
eq.(\ref{split}) is finite for $m_f \rightarrow 0$, so that the $u,d,s$ masses
can be neglected when evaluating this term. $\Delta \alpha$ can be split
into two parts: the contribution from the $e,\mu,\tau$ leptons and the
contribution from the light quarks (i.e. all except $t$), $\Delta
\alpha = \Delta \alpha_{\rm lept} + \Delta \alpha_{\rm had}$.  $\Delta
\alpha_{\rm lep}$ has been calculated up to 3--loop order as
\cite{Steinhauser:1998rq}
\begin{equation}
      \Delta \alpha_{\rm lep} = 0.031497687 \, ,
\end{equation}
while $\Delta \alpha_{\rm had}$ can be extracted from experiment via a
dispersion relation \cite{Hagiwara:2011af}:
\begin{equation}
      \Delta \alpha_{\rm had} = 0.027626\pm0.00138.
\end{equation}
This leads us to 
\begin{eqnarray}
\delta Z_e^{(0)} & = & \frac{1}{2}\frac{\partial}{\partial q^2}
\left[ \Sigma_{\gamma \gamma}^{\rm all \ loops} (q^2)\displaystyle
  |_{q^2=0} -
\Sigma_{\gamma \gamma}^{{\rm light} \ f {\rm \ in \ loops}}
(q^2)\displaystyle |_{q^2=0} \right] \\ 
&  &  +\frac{1}{2}\Delta\alpha + \frac{1}{2 M^2_{Z}} \Re
\Sigma_{\gamma\gamma}^{\textrm{light~} f\textrm{~in loops}}(M^2_{Z}) + \tan\theta_W
\frac{\Sigma_{\gamma Z}(0)}{M^2_{Z}}. 
\end{eqnarray}

One can achieve even higher accuracy by re--summing the light fermion
correction. This corresponds to using the ``running on--shell''
coupling $\alpha(M^2_{Z}) = \alpha(0)/(1-\Delta\alpha) = 1/128.93$ as
input (``tree--level'') coupling.\footnote{As mentioned in Sec.~3,
  this differs slightly from the running $\overline{\rm MS}$ coupling
  $\hat{\alpha}(M_Z) = 1/127.944$ often used in tree--level
  calculations of electroweak processes; see e.g. the review by Erler
  and Langacker in \cite{PDG}.} The corresponding renormalization
constant becomes:
\begin{eqnarray}
e^{\rm bare} \rightarrow e(0)\left(1+\delta Z_e^{(0)}\right) & = & 
e(0)\left(1+\frac{1}{2}\Delta\alpha-\frac{1}{2}\Delta\alpha\right)\left(1+\delta
  Z_e^{(0)}\right)\nonumber\\ 
& = & e(M^2_{Z})\left(1+\delta Z_e^{e(M^2_{Z})} + \rm{higher\,
    order}\right) \, .
\end{eqnarray}
Thus, 
\begin{eqnarray} \label{delta_Ze}
\delta Z_e^{e(M_Z^2)} & = & \delta Z_e^{(0)} -\frac{\Delta \alpha}{2}\nonumber\\
& = & \frac {1}{2} \frac {\partial}{\partial q^2}
\Sigma_{\gamma\gamma}^{\rm all\, loops}(q^2)|_{q^2=0} - \frac {1}{2}
\frac {\partial}{\partial q^2} \Sigma_{\gamma\gamma}^{{\rm light\,} f\,{\rm
  in \, loops}}(q^2)|_{q^2=0} \nonumber \\  
& + &  \frac{1}{2M^2_{Z}} \Re \Sigma_{\gamma\gamma}^{{\rm light\,} f\, {\rm in \,
  loops}}(M^2_{Z}) + \tan\theta_W \frac {\Sigma_{\gamma Z}(0)}
{M^2_{Z}} \, .
\end{eqnarray}

We need $\delta Z_e^{e(M_Z^2)}$ only in the calculation of effective
couplings, hence only the matter (s)fermion loop contributions are included
here.

\section{Higgs Sector}
\label{Appendix:tan}

The only relevant counterterm determined from from this sector 
is $\delta \tan \beta$. It is fixed by forbidding transitions between the
CP--odd Higgs boson $A$ and the $Z$ boson on the $A$ mass shell
\cite{Chankowski:1992er, Dabelstein:1994hb, Dabelstein:1995js}. This gives:
\beq \label{dtb}
\delta\tan \beta = \frac{1}{2 M_Z \cos^{2} \beta }
\Im(\Sigma_{AZ}(m_{A}^{2}))\,.
\eeq
Note that the couplings of $A$ contain an extra factor of imaginary $i$
relative to the gauge couplings. Therefore the imaginary part of the two--point
function appears in eq.(\ref{dtb}); this contains the real (dispersive),
infinite, part of the loop function.

The counterterms for $\cos\beta$ and $\sin\beta$ are thus:
\begin{equation} 
\delta\cos\beta = - \cos^3\beta \tan\beta \delta\tan\beta, 
\end{equation}
\begin{equation} 
\delta\sin\beta = \cos^3\beta \delta\tan\beta. 
\end{equation}

The fermion loop contribution to $\Sigma_{AZ}$ is proportional to the mass of
the fermion in the loop. In case of SM quarks and leptons, this gives an
additional factor of the corresponding Yukawa coupling times $M_W/g$. These
contributions to $\delta \tan\beta$ should therefore be counted as ${\cal
  O}(\lambda^2)$, not as ${\cal O}(\lambda g)$, the second factor of $\lambda$
stemming from the coupling of $A$ to the SM fermion in the loop. 

These counterterms are needed both for the calculation of one--loop corrected
neutralino and chargino masses and for the evaluation of the effective
couplings. As in case of the counterterms to the $W$ and $Z$ masses, in the
former case all MSSM loop contributions are included, while in the latter case
only matter (s)fermion loops are included.


\section{Chargino and Neutralino Sector} 
\label{Appendix:chaneu}

In this appendix, we describe the determination of the relevant counterterms
in the chargino--neutralino sector of the MSSM, closely following the
discussion in \cite{Guasch:2002ez,Fritzsche:2002bi, Baro:2009gn,CDKX}.

The one--loop mass eigenstates can be related to the tree level mass
eigenstates through wave function renormalization:
\beq \label{eZ}
\tilde{\chi_{i}}^{\rm bare} = (\delta_{ij}+\frac{1}{2} \delta Z^L_{ij} P_{L} +
\frac{1}{2} \delta Z^R_{ij} P_{R})~ \tilde{\chi_{j}}^{\rm renormalized}\,.
\eeq 
where $P_L$ and $P_R$ are the chiral projectors. This relation holds
for both the chargino sector, with $\tilde \chi_i \equiv \tilde \chi_i^+, \, i
\in \{1,2\}$, and the neutralino sector, with $\tilde \chi_i \equiv \tilde
\chi_i^0, \, i \in \{1,2,3,4\}$.  The entries of the mixing matrices $N, U$
and $V$ have been chosen to be identical to their tree level values
\cite{Baro:2009gn}.

The corresponding masses receive explicit corrections from one--loop diagrams,
which can be written as
\beq \label{dmf}
\delta m_{f} = \frac{1}{2} m_{f} \left[ \Re\Sigma^{VL}_{ff}(m_{f}^{2}) +\Re 
  \Sigma^{VR}_{ff}(m_{f}^{2}) \right] + \frac{1}{2} \left[ \Re
  \Sigma^{SL}_{ff}(m_{f}^{2}) + \Re \Sigma^{SR}_{ff}(m_{f}^{2}) \right], 
\eeq
where $f$ denotes a fermion species with mass $m_f$. $\Re$ again denotes the
real parts of the loop integrals involved and the $\Sigma$ refer to various
terms in the general two--point function connecting fermions $f$ and $f'$ in
momentum space:
\beq \label{Sigma}
\Sigma_{ff'}(p) = \pslash \left[ P_L \Sigma^{VL}_{ff'}(p) + P_R
  \Sigma^{VR}_{ff'}(p) \right] + P_L \Sigma^{SL}_{ff'}(p) + 
   P_R \Sigma^{SR}_{ff'}(p)\,.
\eeq
Only the case $f=f'$ is needed in eq.(\ref{dmf}), but we will need
off--diagonal two--point functions for the determination of the wave function
counterterms. 

The corrections of eq.(\ref{dmf}) are in general divergent. These divergencies
are absorbed into counterterms to the mass matrices $M^{\rm c}$ and $M^{\rm
  n}$ of the charginos and the neutralinos respectively, i.e.
\beq \label{delM}
M^{\rm bare} = M^{\rm renormalized} + \delta M. 
\eeq
As usual, we define the physical (on--shell) masses as poles of the
real parts of the (one--loop corrected) propagators. The physical
chargino masses are then given by,
\beq \label{m_pchar}
m^{\rm os}_{\tilde\chi_{i}^{+}} = m_{\tilde\chi_{i}^{+}} + (U^* \delta
M^{\rm c}V^{-1})_{ii} - \delta m_{\tilde\chi_{i}^{+}}\,;
\eeq
the corresponding expression for the neutralinos is 
\beq \label{m_pneut}
m^{\rm os}_{\tilde\chi_{i}^{0}} = m_{\tilde\chi_{i}^{0}} + (N^*\delta
M^{\rm n}N^{-1})_{ii} - \delta m_{\tilde\chi_{i}^{0}} \,.
\eeq
The masses $m_{\tilde\chi_i^{+,0}}$ appearing on the right--hand sides
of eqs.(\ref{m_pchar}) and (\ref{m_pneut}) are the (finite)
tree--level masses. $\delta m_{\tilde \chi_i^{+,0}}$ are the explicit 
loop corrections of eq.(\ref{dmf}) as applied to the charginos and neutralinos. 
Finally, $\delta M^{\rm c, n}$ are the counterterm matrices of eq.(\ref{delM}) 
for the chargino and neutralino sector, which we yet have to determine.

To one--loop order counterterms to products like $M_W \sin\beta$ can 
be written as $(\delta M_W) \sin\beta + M_W \delta \sin\beta$, and so on. 

Altogether, the tree--level chargino and neutralino mass matrices depend on
seven parameters, hence there are seven different counterterms: $\delta M_W,\,
\delta M_Z,\, \delta \theta_W, \, \delta \tan\beta, \, \delta M_1$, $\delta
M_2$ and $\delta \mu$. The first three of these already appear in the SM, see
Appendix A. $\delta \tan\beta$ has been fixed in the Higgs sector, see
Appendix \ref{Appendix:tan}. 

Hence only the counterterms to $M_1, \, M_2$ and $\mu$ remain to be
fixed in the chargino and neutralino sector. This means that we can
only chose three of the six masses in this sector to be input masses
which are not changed by loop corrections, i.e. only three of the six
``tree--level'' masses are physical (all--order) masses. We choose the
wino--like chargino mass, the bino--like and higgsino--like neutralino
masses as inputs (i.e. physical masses). Ref.\cite{CDKX} explains in
detail why this particular choice of input states leads to the most
stable perturbative expansion, by avoiding spurious large
corrections. The other three masses will receive finite, but non--zero
corrections. The remaining counterterms can then be computed from
eqs.(\ref{m_pchar}) and (\ref{m_pneut}):
\bea \delta M_{1} & =
&-\dfrac{N_{M_1}}{D},\\ \delta M_{2} & = &\dfrac{N_{M_2}}{D},\\ \delta \mu & =
& \dfrac{N_{\mu}}{D}, \eea
where
\beq \label{D2}
D =  2 N^{*}_{i3} N^{*}_{i4} N^{*^2}_{j1} U^{*}_{k1} V^{*}_{k1}-2
N^{*^2}_{i1} N^*_{j3} N^*_{j4} U^*_{k1} V^*_{k1}
+N^{*^2}_{i2} N^{*^2}_{j1} U^{*}_{k2} V^*_{k2}- N^{*^2}_{i1}
N^{*^2}_{j2} U^{*}_{k2} V^{*}_{k2}, 
\eeq
\bea
N_{\mu}  &=& -\left[N^{*^2}_{j1} \left(\delta m_{\tilde\chi^0_i}-2 \delta M^n_{13}
  N^*_{i1} N^*_{i3}- 2 \delta M^n_{23} N^*_{i2} N^*_{i3} -2
  \delta M^n_{14} N^*_{i1} N^*_{i4} - 2 \delta M^n_{24} N^*_{i2} N^*_{i4} \right)
  \right. \nonumber \\  & & \left.
+ N^{*^2}_{i1} \left(-\delta m_{\tilde\chi^0_j}+ 2 \delta M^n_{13} N^*_{j1} N^*_{j3} 
+ 2 \delta M^n_{23} N^*_{j2} N^*_{j3} + 2 \delta M^n_{14} N^*_{j1} N^*_{j4} +
2 \delta M^n_{24} N^*_{j2} N^*_{j4} \right) \right] U^*_{k1} V^*_{k1} \nonumber \\
&+& \left(-N^{*^2}_{i2} N^{*^2}_{j1}+N^{*^2}_{i1} N^{*^2}_{j2}\right)
\left(-\delta m_{\tilde\chi^+_k}+\delta M^c_{21} U^*_{k2} V^*_{k1}+
\delta M^c_{12} U^*_{k1} V^*_{k2}\right),
\eea
\bea
N_{M_2} & = & 2 \delta m_{\tilde\chi^+_k} N^*_{i3} N^*_{i4}
N^{*^2}_{j1}-2 \delta m_{\tilde\chi^+_k} N^{*^2}_{i1} N^*_{j3}
N^*_{j4} -2 \delta M^c_{21} N^*_{i3}
N^*_{i4} N^{*^2}_{j1} U^*_{k2} V^*_{k1} \nonumber \\
& + & 2 \delta M^c_{21} N^{*^2}_{i1} N^*_{j3} N^*_{j4}
U^*_{k2} V^*_{k1}-2 \delta M^c_{12} N^*_{i3} N^*_{i4}
N^{*^2}_{j1} U^*_{k1} V^*_{k2}-\delta m_{\tilde\chi^0_j}
N^{*^2}_{i1} U^*_{k2} V^*_{k2}\nonumber \\ & + &
 \delta M^c_{12} N^{*^2}_{i1} N^*_{j3} N^*_{j4} U^*_{k1}
 V^*_{k2}+\delta m_{\tilde\chi^0_i} N^{*^2}_{j1} U^*_{k2}
 V^*_{k2} -2 \delta M^n_{13} N^*_{i1} N^*_{i3} N^{*^2}_{j1}
 U^*_{k2} V^*_{k2} \nonumber \\ 
& - & 2 \delta M^n_{23} N^*_{i2} N^*_{i3} N^{*^2}_{j1} U^*_{k2}
 V^*_{k2}-2 \delta M^n_{14} N^*_{i1} N^*_{i4} N^{*^2}_{j1} U^*_{k2}
 V^*_{k2}-2 \delta M^n_{24} N^*_{i2} N^*_{i4} N^{*^2}_{j1}
 U^*_{k2} V^*_{k2}\nonumber \\ 
& + &2 \delta M^n_{13} N^{*^2}_{i1} N^*_{j1} N^*_{j3}
 U^*_{k2} V^*_{k2}+2 \delta M^n_{23} N^{*^2}_{i1} N^*_{j2} N^*_{j3}
U^*_{k2} V^*_{k2}+2 \delta M^n_{14} N^{*^2}_{i1} N^*_{j1}
N^*_{j4} U^*_{k2} V^*_{k2}\nonumber \\ & + &2 \delta M^n_{24}
N^{*^2}_{i1} N^*_{j2} N^*_{j4} U^*_{k2} V^*_{k2},
\eea
\bea
N_{M_1}& = & 2 \delta m_{\tilde\chi^+_k} N^*_{i3} N^*_{i4}
N_{j2}^{*^2}-2 \delta m_{\tilde\chi^+_k} N_{i2}^{*^2} N^*_{j3}
N^*_{j4} -2 \delta m_{\tilde\chi^0_j} N^*_{i3} N^*_{i4}
U^*_{k1} V^*_{k1}-\delta m_{\tilde\chi^0_j} N_{i2}^{*^2}
U^*_{k2} V^*_{k2} 
\nonumber \\ 
& - & 2 \delta M^n_{14} N^*_{i1} N^*_{i4} N_{j2}^{*^2}
U^*_{k2} V^*_{k2}-2 \delta M^n_{24} N^*_{i2} N^*_{i4}
N_{j2}^{*^2} U^*_{k2} V^*_{k2} + 2 \delta M^n_{13} N_{i2}^{*^2}
N^*_{j1} N^*_{j3} U^*_{k2} V^*_{k2} 
\nonumber \\ 
& + & 2 \delta M^n_{23} N_{i2}^{*^2} N^*_{j2} N^*_{j3}
U^*_{k2} V^*_{k2}+2 \delta M^n_{14} N_{i2}^{*^2} N^*_{j1}
N^*_{j4} U^*_{k2} V^*_{k2} + 2 \delta M^n_{24} N_{i2}^{*^2}
N^*_{j2} N^*_{j4} U^*_{k2} V^*_{k2} 
\nonumber \\
&- & 2 \delta M^c_{12} N^*_{i3} N^*_{i4} N_{j2}^{*^2} U^*_{k1}
V^*_{k2}+2 \delta M^c_{12} N_{i2}^{*^2} N^*_{j3} N^*_{j4}
U^*_{k1} V^*_{k2} + 4 \delta M^n_{13} N^*_{i3} N^*_{i4}
N^*_{j1} N^*_{j3} U^*_{k1} V^*_{k1}
\nonumber \\ 
& + & 4 \delta M^n_{23} N^*_{i3} N^*_{i4} N^*_{j2} N^*_{j3}
U^*_{k1} V^*_{k1}-2 \delta M^n_{13} N^*_{i1} N^*_{i3}
N_{j2}^{*^2} U^*_{k2} V^*_{k2} - 2 \delta M^n_{23} N^*_{i2}
N^*_{i3} N_{j2}^{*^2} U^*_{k2} V^*_{k2} 
\nonumber \\
& + &
4 \delta M^n_{14} N^*_{i3} N^*_{i4} N^*_{j1} N^*_{j4}
U^*_{k1} V^*_{k1}+4 \delta M^n_{24} N^*_{i3} N^*_{i4}
N^*_{j2} N^*_{j4} U^*_{k1} V^*_{k1} + 2 \delta
m_{\tilde\chi^0_i} N^*_{j3} N^*_{j4} U^*_{k1} V^*_{k1} 
\nonumber \\
&- & 4 \delta M^n_{24} N^*_{i2} N^*_{i4} N^*_{j3} N^*_{j4}
U^*_{k1} V^*_{k1}-2 \delta M^c_{21} N^*_{i3} N^*_{i4}
N_{j2}^{*^2} U^*_{k2} V^*_{k1} + 2 \delta M^c_{21} N_{i2}^{*^2}
N^*_{j3} N^*_{j4} U^*_{k2} V^*_{k1} 
\nonumber \\
& - & 4 \delta M^n_{13} N^*_{i1} N^*_{i3} N^*_{j3} N^*_{j4}
U^*_{k1} V^*_{k1}-4 \delta M^n_{23} N^*_{i2} N^*_{i3}
N^*_{j3} N^*_{j4} U^*_{k1} V^*_{k1}-4 \delta M^n_{14}
N^*_{i1} N^*_{i4} N^*_{j3} N^*_{j4} U^*_{k1} V^*_{k1} 
\nonumber \\
& + &\delta m_{\tilde\chi^0_i} N_{j2}^{*^2} U^*_{k2} V^*_{k2}.
\eea
Here $i$ and $j$ are the indices of the two neutralino input states, and $k$
is the index of the chargino input state; the counterterms are symmetric under
$i \leftrightarrow j$, both $D$ and the numerators being antisymmetric. Since
we take input states with fixed physical characteristics (bino--, wino-- or
higgsino--like), the numerical values of $i,j,k$ will depend on (the ordering
of) the parameters $M_1, \, M_2$ and $\mu$; see ref.\cite{CDKX} for further
details.

These counterterms are needed both for the calculation of the one--loop
corrected chargino and neutralino masses, where all MSSM loop contributions
have been included in the evaluation of eq.(\ref{dmf}), and for the
calculation of effective couplings (see below), where only matter (s)fermion
loop contributions have been included. 

The diagonal wave--function counterterms are determined by normalizing the
residues of the propagators \cite{Fritzsche:2002bi,Guasch:2002ez}:
\begin{equation} \label{eqzdiag}
\delta Z^{L,R}_i =-
\Re \Sigma^{VL,VR}_{ii}(m_{\tilde\chi_i}^2) - 
m_{\tilde\chi_i}^2\left(\Re \Sigma^{VL \prime}_{ii} 
(m_{\tilde\chi_i}^2)+\Re \Sigma^{VR \prime}_{ii} (m_{\tilde\chi_i}^2)\right)
-2 m_{\tilde\chi_i} \Re \Sigma^{SL,SR \prime}_{ii}(m_{\tilde\chi_i}^2).
\end{equation}
The prime denotes derivative with respect to the external momentum squared.
This expression holds both for charginos [where these counterterms had an
additional superscript $+$ in eqs.(\ref{eq:dUVN2})] and neutralinos.

The off--diagonal wave--function renormalization counterterms are
determined by demanding that the corresponding off--diagonal
two--point correlation functions vanish for on--shell external momenta
\cite{Fritzsche:2002bi, Guasch:2002ez}. This gives:\footnote{Note that
  our $\Sigma_{ij}(m_{\tilde\chi}^2)$ corresponds to
  $\Sigma^{ji}(m_{\tilde\chi}^2)$ of \texttt{FormCalc}.}
\begin{equation} \label{eqzl}
\delta Z^L_{ij} =  \dfrac {2\left(m_{\tilde\chi_i} 
  \hat{\Sigma}_{ij}^{SL}(m_{\tilde\chi_j}^2) + m_{\tilde\chi_j}
  \hat{\Sigma}_{ij}^{SR}(m_{\tilde\chi_j}^2) + m_{\tilde\chi_i}
  m_{\tilde\chi_j} \hat{\Sigma}_{ij}^{VR}(m_{\tilde\chi_j}^2) +
  m_{\tilde\chi_j}^2 \hat{\Sigma}_{ij}^{VL}(m_{\tilde\chi_j}^2)\right)}
{m_{\tilde\chi_i}^2 - m_{\tilde\chi_j}^2},
\end{equation}
\begin{equation} \label{eqzr}
\delta Z^{R}_{ij} = \frac
{2\left(m_{\tilde\chi_i} \hat{\Sigma}_{ij}^{SR}(m_{\tilde\chi_j}^2) 
  + m_{\tilde\chi_j}\hat{\Sigma}_{ij}^{SL}(m_{\tilde\chi_j}^2) + 
  m_{\tilde\chi_i} m_{\tilde\chi_j} \hat{\Sigma}_{ij}^{VL}(m_{\tilde\chi_j}^2) 
  + m_{\tilde\chi_j}^2 \hat{\Sigma}_{ij}^{VR}(m_{\tilde\chi_j}^2)\right)}
 {m_{\tilde\chi_i}^2 - m_{\tilde\chi_j}^2},
\end{equation}
where, 
\begin{equation}
\hat\Sigma^{VL,VR}_{ij} = \Re\Sigma^{VL,VR}_{ij}, \qquad 
\hat{\Sigma}^{SL}_{ij} = \Re\Sigma^{SL}_{ij} - \delta \tilde{m}_{ij}, \qquad 
\hat{\Sigma}^{SR}_{ij} = \Re\Sigma^{SR}_{ij} - \delta \tilde{m}_{ji}^{*}.
\label{renSE}
\end{equation}
These expressions again hold for both charginos and neutralinos, where the
latter were labeled with an additional superscript $+$ in
eqs.(\ref{eq:dUVN2}) in the main text.  Here $\delta \tilde{m}_{ij}$ denotes
$(U^{*}\delta M^{c} V^{-1})_{ij}$ and $(N^{*}\delta M^{n} N^{-1})_{ij}$ for
charginos and neutralinos, respectively. The calculation of the counterterm
matrices $\delta M^c$ and $\delta M^n$ has been discussed in the first part of
this Appendix. Hermiticity implies the following relations:
$$ 
\Sigma^{SR}_{ij}(q^2) = \Sigma^{SL*}_{ji}(q^2),~
\Sigma^{SL}_{ij}(q^2) = \Sigma^{SR*}_{ji}(q^2),~
\Sigma^{VL}_{ij}(q^2) = \Sigma^{VL*}_{ji}(q^2),~
\Sigma^{VR}_{ij}(q^2) = \Sigma^{VR*}_{ji}(q^2).
$$ 
Neutralinos are Majorana fermions. Consequently the following additional
relations \cite{Fritzsche:2002bi,Baro:2009gn} hold:
$$
\Sigma^{RS}_{ij}(q^2) = \Sigma^{RS}_{ji}(q^2),~
\Sigma^{LS}_{ij}(q^2) = \Sigma^{LS}_{ji}(q^2),~
\Sigma^{LV}_{ij}(q^2) = \Sigma^{RV}_{ji}(q^2).
$$
These relations can be used to further simplify eqs.(\ref{renSE}) as
applied to the neutralinos \cite{Fritzsche:2002bi,Baro:2009gn}. 

Note finally that neutralino couplings may be purely imaginary even if
CP is conserved; this is true for states corresponding to negative
eigenvalues of the neutralino mass matrix. Since $\Re$ {\em only} acts
on the loop functions, these imaginary parts have to be kept in
eqs.(\ref{eqzdiag})--(\ref{eqzr}); in fact, they are necessary to
yield finite results in eqs.(\ref{eqzl}), (\ref{eqzr}) for
$m_{\tilde\chi_i^0} = m_{\tilde \chi_j^0}$, which is possible if $M_1
M_2 < 0$. In this case $\tilde \chi_i^0$ and $\tilde \chi_j^0$ can be
combined into a Dirac neutralino state, and the corresponding
``off--diagonal'' wave function renormalization becomes part of the
diagonal wave function renormalization of this Dirac neutralino.
Degenerate charginos can only occur if $\tan\beta = 1$, which is
excluded by Higgs mass bounds, or in the presence of nontrivial
CP--violating phases in the chargino mass matrix.

\bibliographystyle{utphys}
\addcontentsline{toc}{section}{References}
\bibliography{cdk}
\end{document}